\documentclass[final,authoryear,5p,times,twocolumn]{elsarticle}
\usepackage{epsfig}

\usepackage{amssymb}
\usepackage{lineno}
\usepackage{rotating}
\usepackage{color,soul}

\usepackage[bookmarks = true, bookmarksnumbered = true, pdfpagemode =None, pdfstartview = FitH, pdfpagelayout = SinglePage, colorlinks = true, urlcolor = blue, citecolor = monbleu]{hyperref}

\newcommand{\pderiv}[2]{\frac{\partial #1}{\partial #2}}

\newcommand{\beq}{\bigskip\begin{equation}}
\newcommand{\eeq}{\bigskip\end{equation}}

\journal{Icarus}

\begin{document}

\begin{frontmatter}



\title{Seasonal Variability of Saturn's Tropospheric Temperatures, Winds and Para-H$_2$ from Cassini Far-IR Spectroscopy}


\author[le,ox]{Leigh N. Fletcher}
\ead{leigh.fletcher@leicester.ac.uk}
%\author[swri]{T.K. Greathouse}
\author[ox]{P.G.J. Irwin}
%\author[om]{O. Mousis}
%\author[ox]{J.A. Sinclair}
\author[umd]{R.K. Achterberg}
\author[jpl]{G.S. Orton}
\author[gsfc]{F.M. Flasar}

%\author[ox]{R.S. Giles}
%\author[ral]{J. Hurley}
%\author[cu]{N. Gorius}
%
%\author[umd]{B.E. Hesman}
%\author[gsfc]{G.L. Bjoraker}

\address[le]{Department of Physics \& Astronomy, University of Leicester, University Road, Leicester, LE1 7RH, UK}
\address[ox]{Atmospheric, Oceanic \& Planetary Physics, Department of Physics, University of Oxford, Clarendon Laboratory, Parks Road, Oxford, OX1 3PU, UK}
%\address[swri]{Southwest Research Institute, Division 15, 6220 Culebra Road, San Antonio, Texas 78228, USA}
\address[umd]{Department of Astronomy, University of Maryland, College Park, MD 20742, USA.}
\address[jpl]{Jet Propulsion Laboratory, California Institute of Technology, 4800 Oak Grove Drive, Pasadena, CA, 91109, USA}
%\address[om]{Universit\'{e} de Franche-Comt\'{e}, Institut UTINAM, CNRS/INSU, UMR 6213, Besan\c{c}on, Cedex, France}
%\address[ral]{STFC Rutherford Appleton Laboratory, Harwell Science and Innovation Campus, Didcot, OX11 0QX, UK}
%\address[cu]{Department of Physics, The Catholic University of America, Washington, DC 20064, USA}
%\address[umd]{Department of Astronomy, University of Maryland, College Park, MD 20742, USA.}
\address[gsfc]{NASA/Goddard Space Flight Center, Greenbelt, Maryland, 20771, USA }

%\linenumbers

\begin{abstract}

Far-IR 16-1000 $\mu$m spectra of Saturn's hydrogen-helium continuum measured by Cassini's Composite Infrared Spectrometer (CIRS) are inverted to construct a near-continuous record of upper tropospheric (70-700 mbar) temperatures and para-H$_2$ fraction as a function of latitude, pressure and time for a third of a Saturnian year (2004-2014, from northern winter to northern spring).  The thermal field reveals evidence of reversing summertime asymmetries superimposed onto the belt/zone structure.  The temperature structure that is almost symmetric about the equator by 2014, with seasonal lag times that increase with depth and are qualitatively consistent with radiative climate models. Localised heating of the tropospheric hazes (100-250 mbar) create a distinct perturbation to the temperature profile that shifts in magnitude and location, declining in the autumn hemisphere and growing in the spring.  Changes in the para-H$_2$ ($f_p$) distribution are subtle, with a 0.02-0.03 rise over the spring hemisphere (200-500 mbar) perturbed by (i) low-$f_p$ air advected by both the springtime storm of 2010 and equatorial upwelling; and (ii) subsidence of high-$f_p$ air at northern high latitudes, responsible for a developing north-south asymmetry in $f_p$.  Conversely, the shifting asymmetry in the para-H$_2$ disequilibrium primarily reflects the changing temperature structure (and hence the equilibrium distribution of $f_p$), rather than actual changes in $f_p$ induced by chemical conversion or transport.  CIRS results interpolated to the same point in the seasonal cycle as re-analysed Voyager-1 observations (early northern spring) show qualitative consistency from year to year (i.e., the same tropospheric asymmetries in temperature and $f_p$), with the exception of the tropical tropopause near the equatorial zones and belts, where downward propagation of a cool temperature anomaly associated with Saturn's stratospheric oscillation could potentially perturb tropopause temperatures, para-H$_2$ and winds.  Quantitative differences between the Cassini and Voyager epochs suggest that the oscillation is not in phase with the seasonal cycle at these tropospheric depths (i.e., it should be described as quasi-periodic rather than `semi annual').  Variability in the zonal wind field derived from latitudinal thermal gradients is small ($<10$ m/s per scale height near the tropopause) and mostly affects the broad retrograde jets, with the notable exception of large variability on the northern flank of the equatorial jet.  The meridional potential vorticity (PV) gradient, and hence the `staircase of PV' associated with spatial variations in the vigour of vertical mixing, has varied over the course of the mission but maintained its overall shape.  PV gradients in latitude and altitude are used to estimate the atmospheric refractive index for the propagation of stationary planetary (Rossby) waves, predicting that such wave activity would be confined to regions of real refractivity (tropical regions plus bands at $35-45^\circ$ in both hemispheres).  The penetration depth of these regions into the upper troposphere is temporally variable (potentially associated with stratification changes), whereas the latitudinal structure is largely unchanged over time (associated with the zonal jet system).  

\end{abstract}


\end{frontmatter}

%\linenumbers

%%%%%%%%%%%%%%%%%%%%%%%%%%%%%%%%%%%%%%%%%%%%%%
%%%%%%%%%%%%%%%%%%%%%%%%%%%%%%%%%%%%%%%%%%%%%%
%%%%%%%%%%%%%%%%%%%%%%%%%%%%%%%%%%%%%%%%%%%%%%
\section{Introduction}
\label{intro}

% Tropospheric temperature review
Far-infrared remote sensing (16-1000 $\mu$m) provides access to the three-dimensional structure of Saturn's upper troposphere between approximately 70 and 700 mbar.  This altitude range lies above the main condensible cloud layers \citep[expected to be NH$_3$ ice below the 1.2-bar level,][]{99atreya}, but contains both the radiative-convective (R-C) boundary and the ubiquitous haze that is responsible for Saturn's yellow-ochre hues.  The R-C boundary signifies the transition from the convectively-unstable mid-troposphere to the statically-stable upper troposphere.  The temperature field in this region is characterised by latitudinal contrasts between the cool zones (anticyclonic shear regions equatorward of prograde jets) and warmer belts (cyclonic shear regions poleward of prograde jets) \citep{83conrath}.  When Cassini first arrived at Saturn in 2004 (just after southern summer solstice at a planetocentric solar longitude of $L_s=293^\circ$), inversions of spectra from the Cassini Composite Infrared Spectrometer \citep[CIRS,][]{04flasar} revealed stark hemispheric contrasts in upper tropospheric temperatures and composition, caused by a combination of radiative, dynamical and chemical responses to the seasonal shift in sunlight \citep{07fletcher_temp}.  Qualitatively similar asymmetries had been observed 24 years earlier by Voyager 1 (1980, $L_s=8.6^\circ$) and Voyager 2 (1981, $L_s=18.2^\circ$) \citep{81hanel, 82hanel, 83conrath, 98conrath}, although a direct quantitative comparison of Cassini and Voyager findings only became possible in 2010-2011 when Saturn was at the same point in its seasonal cycle.

Cassini has continued to study Saturn's evolving atmosphere through spring equinox ($L_s=0^\circ$, August 2009) and on towards northern summer solstice ($L_s=90^\circ$, May 2017).  Several studies have used CIRS 7-16 $\mu$m mid-infrared spectroscopy to monitor the evolution of asymmetries in temperature at pressures less than 250 mbar and stratospheric hydrocarbon composition \citep{10fletcher_seasons, 13sinclair, 15sylvestre, 15fletcher_poles}.  The broad temperature trends compare favourably to the predictions of Saturnian climate models \citep{10greathouse, 12friedson, 14guerlet} that balance atmospheric heating from methane and aerosol absorption with cooling by hydrocarbon emission.  This study utilises far-infrared (16-1000 $\mu$m) CIRS spectroscopy to investigate seasonal trends in the deeper troposphere, constructing a global record of tropospheric temperature and para-hydrogen variability over the ten-year span of the Cassini mission.  
%These results can now be quantitatively compared to the distributions derived by Voyager/IRIS a Saturnian-year earlier.  
%Ultimately, this slow seasonal variability will be compared to the shifting haze distribution observed in reflected sunlight \citep[e.g.,][]{12edgington} to understand the link between haze production and tropospheric conditions.

% Para-Hydrogen Review
Saturn's 16-1000 $\mu$m spectrum is shaped by the collision-induced absorption of H$_2$-H$_2$ and H$_2$-He.  Collisions between these molecules create instantaneous dipoles that form the continuum, and the ratio of ortho-H$_2$ (corresponding to the odd rotational isomer of H$_2$ with parallel spins) to para-H$_2$ (the even isomer of H$_2$ with anti-parallel spins) can be deduced from the relative sizes of the broad S(0) (due to para-H$_2$) and S(1) (due to ortho-H$_2$) absorption features near 354 and 587 cm$^{-1}$, respectively.  In Saturn's deep troposphere the ratio between these two spin isomers is expected to be in equilibrium (3:1), equivalent to the para-hydrogen fraction ($f_p$) of 0.25 \citep[e.g.,][]{82massie}.  CIRS constraints on $f_p$ mostly arise from the S(0) line and the translational continuum at wavenumbers less than 200 cm$^{-1}$, as the signal-to-noise ratio decreases substantially near the S(1) line.  At the cold temperatures of the upper troposphere the equilibrium para-H$_2$ fraction ($f_{eqm}$) increases beyond 0.25 (reaching 0.4-0.5 at the tropopause) because of the increased population of para-H$_2$ energy levels at lower temperatures.  However, if an upward-moving air parcel is rising quickly enough that the vertical mixing is faster than the chemical equilibration timescale (i.e., assuming negligible ortho-para conversion during the motion), then the upwelling parcel will tend to retain its initial low para-H$_2$ fraction.    The para-H$_2$ fraction would then appear to be in disequilibrium in Saturn's upper troposphere.   The spatial distribution of the para-H$_2$ fraction, and its deviation from equilibrium, can therefore be used to trace both vertical mixing and the efficiency of chemical equilibration, with strong upwelling causing sub-equilibrium conditions ($f_p<f_{eqm}$) and subsidence of cool tropopause air causing super-equilibrium conditions ($f_p>f_{eqm}$) \citep{98conrath, 07fletcher_temp}.  

Voyager/IRIS revealed local sub-equilibrium conditions near $60^\circ$S in early northern spring \citep[$L_s=8.6-18.2^\circ$][]{98conrath}, and an asymmetry with a higher $f_p$ in the spring hemisphere (super-equilibrium conditions from 70$^\circ$N to 30$^\circ$S).  Conversely, Cassini observations in late northern winter ($L_s=297-316^\circ$) revealed a region of low $f_p$ at the equator (interpreted as evidence of equatorial upwelling), super-equilibrium over much of the southern summer hemisphere and sub-equilibrium in the winter hemisphere poleward of $30^\circ$N \citep{07fletcher_temp}. The difference between the Voyager and Cassini results was believed to be seasonal, but a continuous Cassini time series (rather than a single snapshot) is required to confirm this, and to distinguish between dynamical and chemical explanations for the observed $f_p$ distribution.  

% Temperature knee
The presence of aerosols may serve to catalyse the interconversion between the ortho- and para-H$_2$ isomers \citep[e.g.,][]{82massie,03fouchet}, so Saturn's upper tropospheric haze is expected to be intimately linked to both the temperature and para-H$_2$ distribution.  The composition of the haze is unknown, possibly linked to uplift of condensed NH$_3$ ice, subsidence of photochemically-produced hydrocarbon smog, or photochemical destruction of tropospheric PH$_3$ to form P$_2$H$_4$ as a haze precursor \citep[see the review by][]{09west}.  The haze appears to be seasonally variable, and responsible for the observed colour change from blue to yellow as Saturn's northern hemisphere emerged from the depths of winter \citep[e.g.,][]{12edgington}.  CIRS identified a local maximum in tropospheric temperatures (a `kink' in the vertical temperature profile) between the tropopause and the radiative-convective boundary \citep{07fletcher_temp}, that had been previously identified at isolated southern latitudes by Voyager \citep{81hanel, 85lindal}.  The pressure level and magnitude of this temperature perturbation was latitudinally variable, being enhanced in southern summer and weak or absent in the northern winter hemisphere.  The perturbation was also higher and weaker over the equator, and showed local maxima at $\pm15^\circ$ latitude coincident with the neighbouring warm equatorial belts.  \citet{07fletcher_temp} concluded that this was a radiative effect due to solar absorption Saturn's upper tropospheric haze, and explained the asymmetry in terms of both seasonal insolation and the variable distributions of tropospheric aerosols \citep[later confirmed by radiative climate modelling by][]{12friedson, 14guerlet}.  This study attempts to track the evolution of the upper tropospheric temperature perturbation in relation to the distribution of hazes.

%Summary
Section \ref{cirs} describes the processing and analysis of Cassini/CIRS far-infrared spectra to reconstruct a continuous record of tropospheric temperatures and para-H$_2$ fraction since 2004 (Section \ref{results}). Section \ref{iris} then compares the results in early northern spring to those derived from Voyager/IRIS observations to identify inter-annual variability, before discussing the implications of this study in Section \ref{discuss}.  We will use the derived temperature and para-hydrogen fields to estimate the atmospheric stratification and zonal wind field in three dimensions (altitude, latitude and time), calculating the implied potential vorticity gradients and their implications for seasonally-variable mixing and the transmissivity of planetary waves in Saturn's upper troposphere.  In particular, measurements of the 2D wind and potential vorticity gradients from Cassini data will provide insight into the dynamical processes occurring on the scale of Saturn's jet system, a measure of the intrinsic stability of the features observed in the flow, and a closer connection between observations and numerical simulations of giant planet atmospheres \citep{06read_jup,09read}.

%%%%%%%%%%%%%%%%%%%%%%%%%%%%%%%%%%%%%%%%%%%%%%
%%%%%%%%%%%%%%%%%%%%%%%%%%%%%%%%%%%%%%%%%%%%%%
%%%%%%%%%%%%%%%%%%%%%%%%%%%%%%%%%%%%%%%%%%%%%%
\section{Data Analysis}
\label{cirs}

\subsection{Cassini/CIRS Data}

\begin{figure}
\begin{centering}
\includegraphics[angle=0,scale=.75]{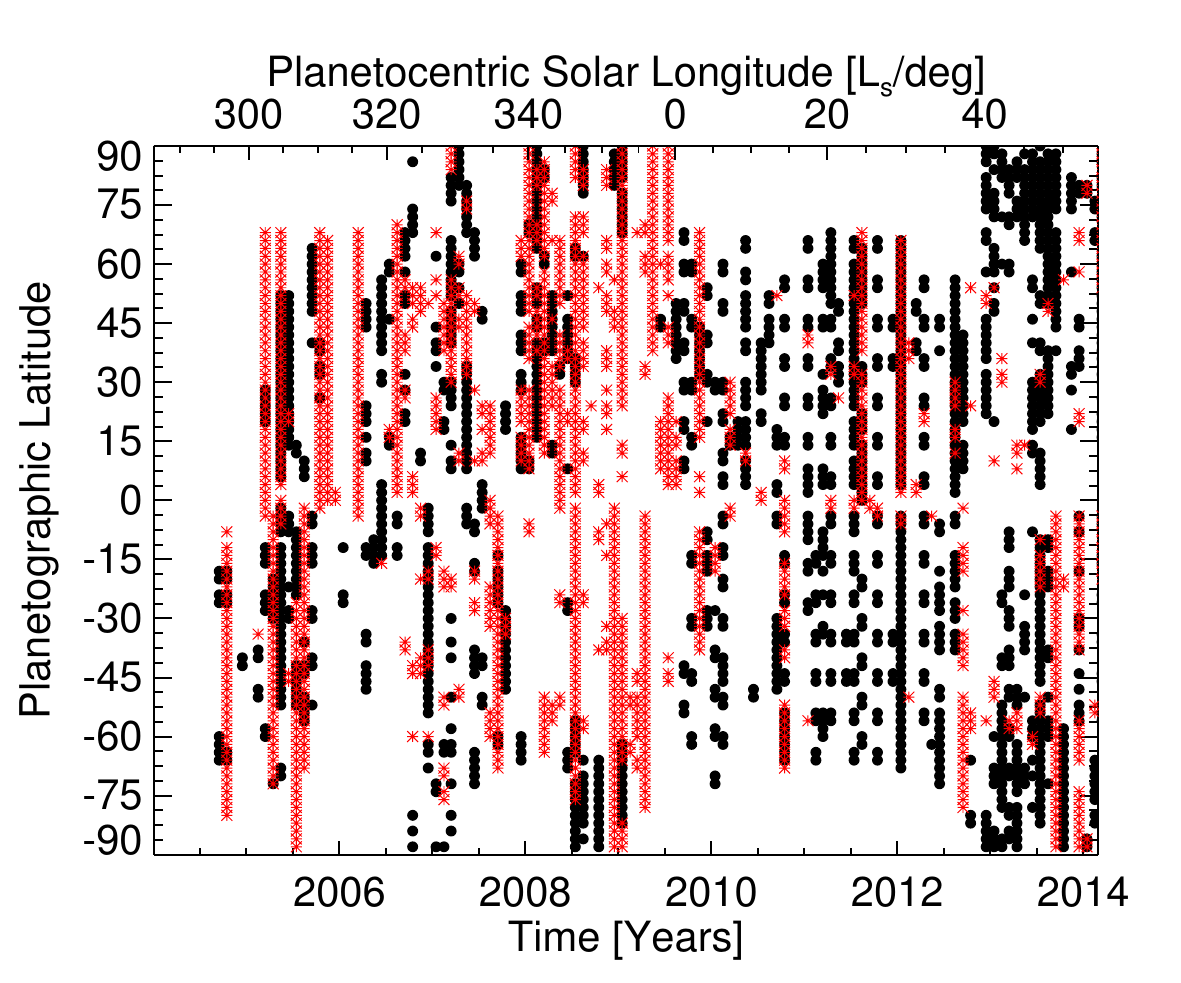}
\caption{Spatial and temporal distribution of more than 3000 CIRS focal plane one footprints at 2.5 cm$^{-1}$ spectral resolution (black dots) and 15.0 cm$^{-1}$ spectral resolution (red crosses).}
\label{coverage}
\end{centering}
\end{figure}

CIRS is made up of two interferometers sharing a common telescope and scan mechanism \citep[instrument and observation strategies are reviewed by][]{04flasar}.  This study primarily uses nadir spectra acquired with the longer wavelength ($\lambda>16$ $\mu$m) polarising interferometer (focal plane 1), measuring the 10-600 cm$^{-1}$ region of Saturn's far-infrared spectrum using a single pixel with a circular field of view (4.1 mrad diameter, with a Gaussian spatial response function of full width at half maximum of 2.5 mrad).  Additional mid-infrared spectra sensitive to stratospheric methane emission (7.7 $\mu$m in focal plane 4) are included to constrain stratospheric temperature.  CIRS has a tuneable spectral resolution between 0.5 and 15 cm$^{-1}$, and we use the 2.5 and 15 cm$^{-1}$ settings.  The identification of trends in zonal mean atmospheric properties requires a significant number of points in a time series, so we follow \citet{15fletcher_poles} and use both `targeted' and `ride-along' observations (i.e., those where CIRS acquires data during observations by other Cassini instruments).  The spatial and temporal distribution of the CIRS spectra used in this study are shown in Fig. \ref{coverage}.  This incorporates both `sit-and-stare' and scanning observations, and increases the temporal coverage beyond the `FIRMAP' datasets used in previous studies of Saturn's temperature field \citep[red crosses in Fig. \ref{coverage}][]{07fletcher_temp, 10fletcher_seasons, 13sinclair}, at the expense of providing incomplete longitudinal coverage at a particular latitude and time.  The spatial resolution of the observations were generally greater for the 15-cm$^{-1}$ data ($2-5^\circ$ latitude) than for the 2.5 cm$^{-1}$ data (4-12$^\circ$), although this was highly variable over Cassini's orbital tour.

All CIRS 2.5- and 15-cm$^{-1}$ nadir spectra from 2004 to 2014 were extracted from the latest version of the CIRS database (version 4.3, Fig. \ref{coverage}) using month-long bins at monthly intervals and a $4^\circ$ latitudinal grid (stepped every $2^\circ$), but applying strong filtering to omit spectra with calibration defects or inappropriate observing geometries, as described by \citet{15fletcher_poles}.  Spectra were coadded within $\pm10^\circ$ of the mean zenith angle, and we required at least 40 target spectra and 4000 deep space calibration spectra for the coadded spectrum to be considered for spectral inversion.  The coadded spectra were assumed to be representative of the zonal means, given that large-scale longitudinal perturbations \citep[e.g., thermal contrasts due to Saturn's northern springtime storm,][]{11fletcher_storm, 14achterberg} are extremely rare.  Note that the storm-perturbed latitudes were omitted between 2011 and 2013 for the purpose of this study. This resulted in 1650 monthly-zonal-mean 2.5-cm$^{-1}$ spectra and 1520 monthly-zonal-mean 15-cm$^{-1}$ spectra between 2004 and 2014.

\begin{figure}
\begin{centering}
\includegraphics[angle=0,scale=.75]{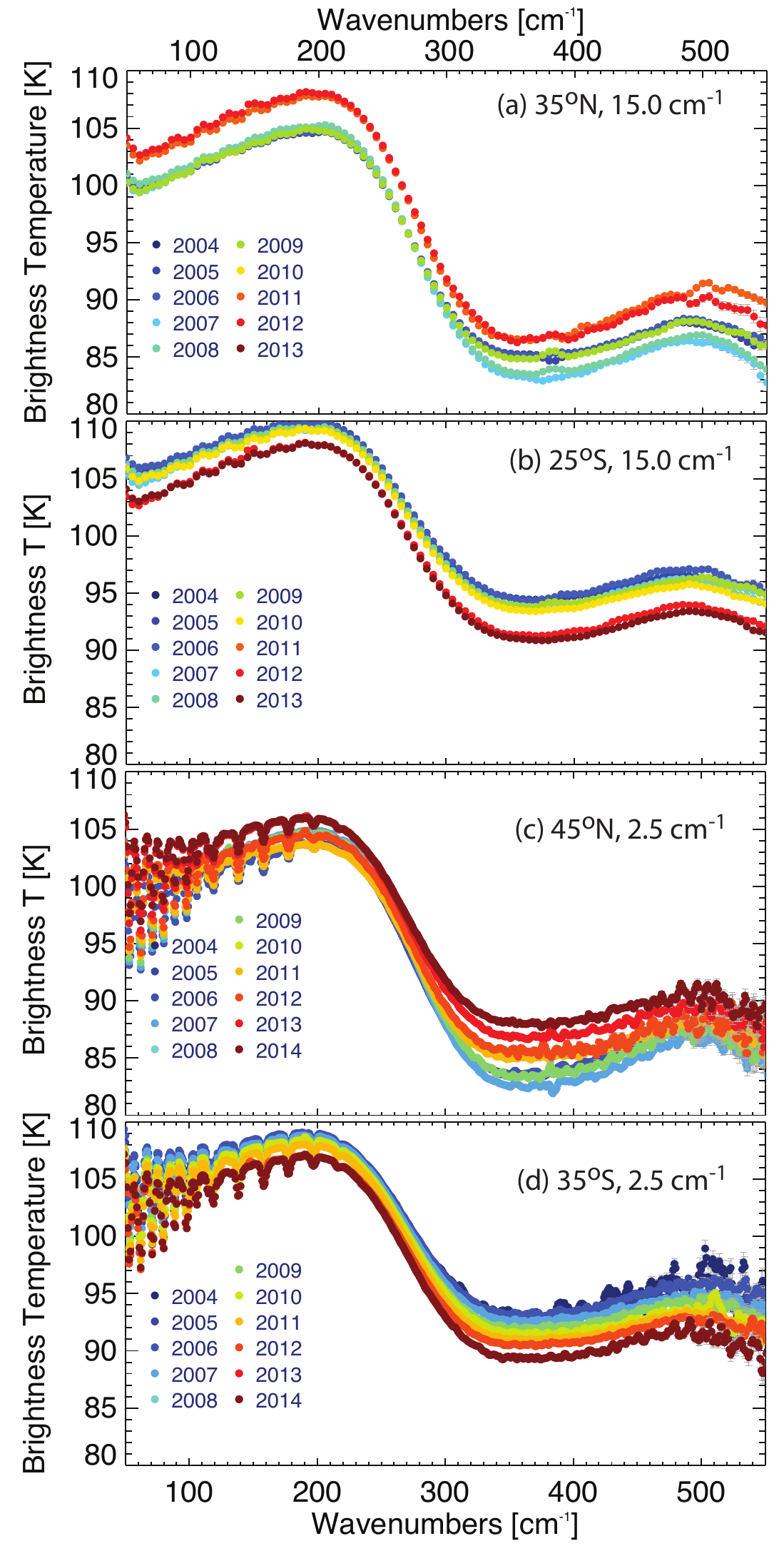}
\caption{Four examples of annually-averaged Cassini/CIRS far-IR (FP1) brightness temperature spectra, showing the difference between 15-cm$^{-1}$ resolution (a, b) and 2.5-cm$^{-1}$ resolution (c, d).  Spectra were averaged over a full year and a $10^\circ$ latitude range, taking care to match emission angles.  Not all latitudes are covered for every year.  Warming trends in the northern hemisphere and cooling trends in the southern hemisphere are evident, as well as subtle shape changes due to altering para-H$_2$.  Error bars are plotted, but are smaller than the individual data points for wavenumbers below 450 cm$^{-1}$. }
\label{spx}
\end{centering}
\end{figure}

Examples of annual-mean far-IR spectra are shown in Fig. \ref{spx}, where for the purposes of display the spectra were coadded over a full year within a $10^\circ$-wide latitude bin and $10^\circ$-wide zenith angle bin.  Rotational lines of PH$_3$ and NH$_3$ are readily visible at 2.5 cm$^{-1}$ (Fig. \ref{spx}c,d), but also cause oscillations of the 50-200 cm$^{-1}$ spectrum at 15 cm$^{-1}$ resolution.  Although formal spectral inversions are required to distinguish the contributions of temperature and the para-H$_2$ fraction, subtle changes in para-H$_2$ are manifest as changes to the spectral slope (e.g., Fig. \ref{spx}c), whereas shifts in temperatures generally scale the spectrum up and down (see Fig. \ref{spxvar}).

\subsection{Spectral Inversion}

% Basics of retrievals
CIRS spectra for every month and latitude were inverted to derive atmospheric temperature and composition using the optimal estimation retrieval algorithm, NEMESIS \citep{08irwin}.  A Levenburg-Marquardt iterative scheme is used to minimise the residual between the forward model and the measurements, whilst maintaining physically-realistic solutions by using smooth \textit{a priori} state vectors \citep[see][for a full discussion of optimal solutions to the ill-posed inverse problem]{00rodgers}.  The \textit{a priori} atmospheric state consists of: (a) a reference temperature profile based on previous CIRS analyses \citep{07fletcher_temp}, which were themselves based on Voyager/IRIS and radio science measurements \citep{84courtin, 85lindal}, defined on 120 pressure levels equally spaced in $\log{p}$ between 1 $\mu$bar and 10 bar; (b) the equilibrium para-H$_2$ distribution; and (c) gaseous profiles for methane \citep[fixed at a mole fraction of $4.7\times10^{-3}$,][]{09fletcher_ch4}, ammonia \citep{12hurley}, phosphine \citep[based on the global mean results of][]{09fletcher_ph3}, and the stratospheric hydrocarbons ethane and acetylene \citep[using a spatial average of results from][]{09guerlet}.  Rotational lines of NH$_3$ and PH$_3$ have a non-negligible influence on the shape of Saturn's spectrum below 220 cm$^{-1}$; stratospheric hydrocarbons form emission features long ward of 650 cm${-1}$, and so are included in our model for consistency only.  

% Reference line data
Sources of spectral line data in our forward model are identical to those in Table 4 of \citet{12fletcher}.  These absorption coefficients are ranked according to their frequency distributions within a small spectral interval, producing smoothly-varying $k$-distributions that increase the efficiency of transmission calculations via the correlated-$k$ approximation \citep{89goody_ck, 91lacis}.  Absorption coefficients for the collision-induced H$_2$-H$_2$ continuum were based on the results of \citet{07orton} (provided by M. Gustafsson for a range of $f_p$ values), who updated the calculations of \citet{85borysow}.  Additional opacities from H$_2$-He, H$_2$-CH$_4$ and CH$_4$-CH$_4$ were included from \citet{87borysow, 86borysow, 88borysow}.  These low-temperature opacities are the same as those reported by \citet{12richard} for $T<400$ K and included in the `Alternative' folder of the HITRAN 2012 compilation \citep{13rothman}.  The synthetic spectra were convolved with a triangular instrument function with a width appropriate to the spectral resolution of the measurements, approximating the Hamming apodisation of CIRS.  Functional derivatives (or Jacobians, the rate of change of spectral radiance with the parameter of interest) calculated for Saturn's reference temperature span 100-800 mbar in the 10-500 cm$^{-1}$ region and 1-3 mbar in the 1250-1350 cm$^{-1}$ region of stratospheric CH$_4$ emission \citep[see Table 2 and Fig. 1 of][]{07fletcher_temp}.   Temperatures are smoothly interpolated between these two regions, and simply relax back to the \textit{a priori} for $p>800$ mbar and $p<1$ mbar.  Para-H$_2$ sensitivity in CIRS far-IR spectra ranges from 100-550 mbar with a peak near 270 mbar, and CIRS cannot constrain $f_p$ at altitudes above the tropopause.

\begin{figure}
\begin{centering}
\includegraphics[angle=0,scale=.90]{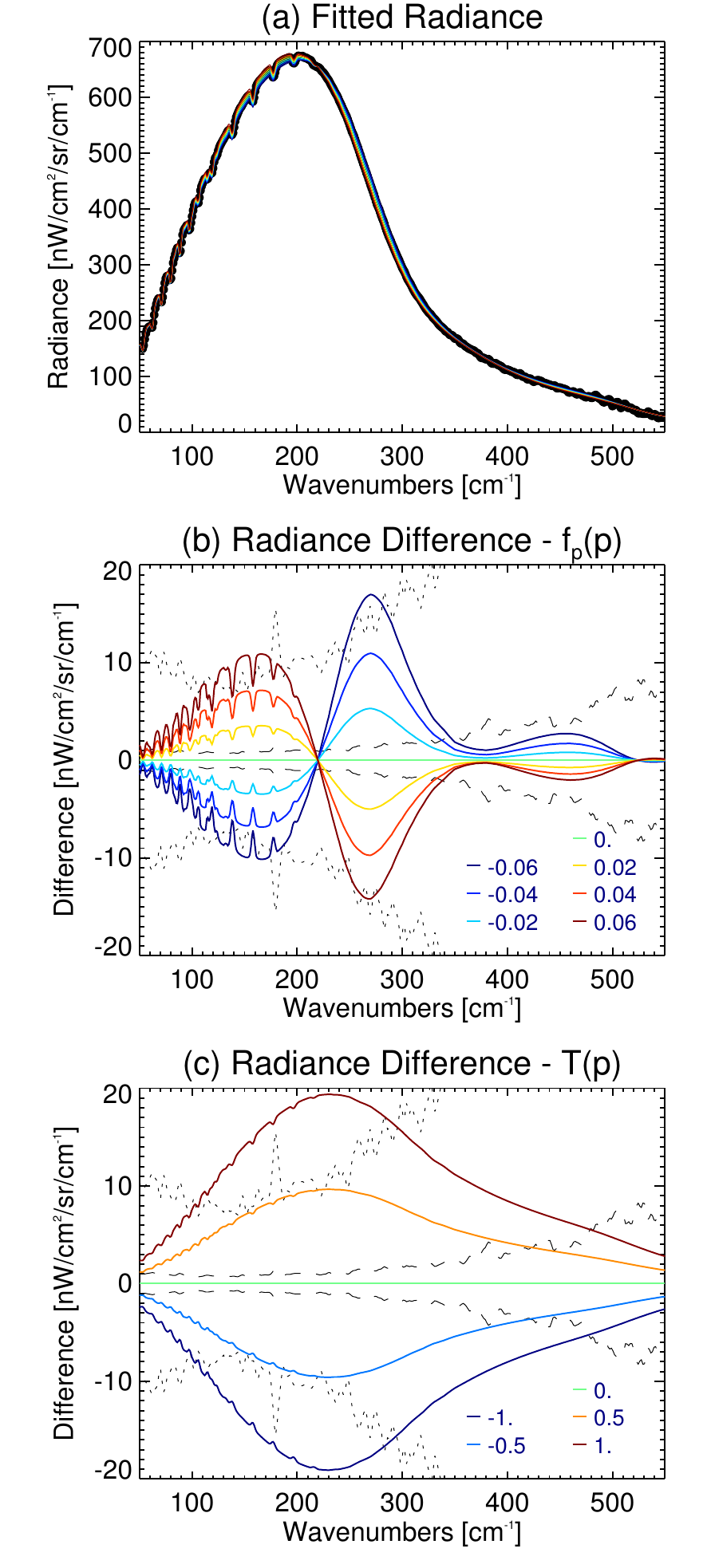}
\caption{Example of the effect on the CIRS spectral radiance of changing either the para-H$_2$ fraction or the tropospheric temperatures.  The 2.5-cm$^{-1}$ spectrum shown as dots in panel a is for $45^\circ$N and 2014, as shown in brightness temperature units in Fig. \ref{spx}c.  The solid lines through these points represent a typical fit to the FP1 spectrum.  Panel a indicates that changes of order 10 nW/cm$^2$/sr/cm$^{-1}$ are barely discernible given the size of the radiances themselves.  However, the difference plot in panel b shows that changing $f_p$ alters the shape of the spectral continuum, whereas panel c shows that changing the temperature alters the overall intensity of the spectrum.  In b and c, the dotted line shows the CIRS noise-equivalent spectral radiance for a single spectrum; the dashed line shows the uncertainty when 100 spectra are coadded (a typical number for this study). }
\label{spxvar}
\end{centering}
\end{figure}

% Retrieval procedure
For each spectrum, we simultaneously retrieved profiles of temperature and $f_p$ (in addition to scale factors for the \textit{a priori} distributions of PH$_3$ and NH$_3$) from the 50-550 cm$^{-1}$ region of CIRS focal plane one and the 1230-1380 cm$^{-1}$ region of focal plane four.  The latter was added to the analysis to ensure that seasonal temperature contrasts in the lower stratosphere \citep{10fletcher_seasons} were being correctly accounted for in the tropospheric retrieval.  \mbox{Fig. \ref{spxvar}} shows why the temperature and para-H$_2$ are separable in the inversion, as they have dramatically different effects on the shape of the collision-induced continuum.  Temperature changes of $\pm1$ K modulate the continuum intensity in \mbox{Fig. \ref{spxvar}c}, whereas changes in the para-H$_2$ fraction of $\pm0.06$ (the largest changes seen in this study) modify the spectral shape in \mbox{Fig. \ref{spxvar}b}.  These differences are easily detectable compared to the noise-equivalent spectral radiance on the coadded CIRS spectra.

Although PH$_3$ and NH$_3$ affect the spectral shape between 50 and 220 cm$^{-1}$, the retrieved mole fractions were found to be positively correlated with the retrieved para-H$_2$ fractions in the upper troposphere because the $f_p$ retrievals use the translational continuum shortward of 200 cm$^{-1}$ in addition to the broad S(0) and S(1) lines (Fig. \ref{spxvar}b).  As a result, large excursions in one parameter would occasionally be mirrored by large excursions in the other.  Rather than simply fixing the gases at their priors, we allowed them to vary during the retrieval but filtered out substantial outliers from the final time series, as described in the next Section.  Quantitative estimates of the para-H$_2$ fraction should be treated with caution, as they are sensitive to the helium mole fraction \citep[e.g.,][]{07fletcher_temp}, which remains poorly constrained on Saturn.  Our study fixes helium at a mole fraction of 0.135 as derived by Voyager/IRIS \citep{00conrath}.  CIRS focal plane 3 (600-1100 cm$^{-1}$), which also samples the upper troposphere, is omitted from this study because the FP1 and FP3 detector footprints don't always overlap and therefore sample different horizontal regions.  Although this is also true of focal plane 4 (used to constrain CH$_4$ emission), FP4 largely samples the stratosphere where spatial variability occurs over broader scales, and therefore has negligible effect on the tropospheric para-H$_2$ retrievals.  Finally, it is interesting to note that a CIRS focal plane 2, which was descoped during a Cassini downsizing, would have provided enhanced signal-to-noise ratio near the S(1) line to improve our ability to constrain Saturn's para-H$_2$ distribution.

%; and (ii) the higher signal-to-noise of the 600-700 cm$^{-1}$ region was driving the para-H$_2$ retrieval at the expense of poorly fitting the 50-550 cm$^{-1}$ region.  

%%%%%%%%%%%%%%%%%%%%%%%%%%%%%%%%%%%%%%%%%%%%%%
%%%%%%%%%%%%%%%%%%%%%%%%%%%%%%%%%%%%%%%%%%%%%%
%%%%%%%%%%%%%%%%%%%%%%%%%%%%%%%%%%%%%%%%%%%%%%
\section{Cassini Results}
\label{results}

\begin{figure}
\begin{centering}
\includegraphics[angle=0,scale=.80]{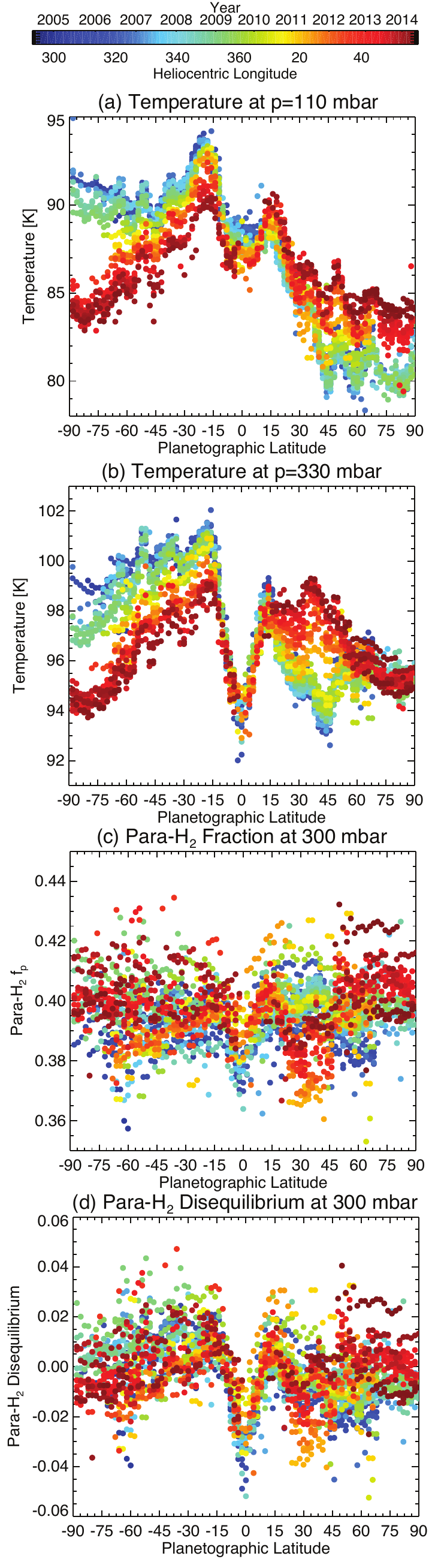}
%\noindent\makebox[\textwidth]{\includegraphics[height=\textheight]{figs/temp_fpara.pdf}}
\caption{Variability of tropospheric temperature and para-H$_2$ over the timespan of Cassini observations.  The colour of each measurement represents the month of observation according to the key at the top of the figure. Thermal changes at 110 and 330 mbar (a,b) exceed the 0.5-1.0 K uncertainty on retrieved temperatures at each altitude.  However, the scatter in retrieved $f_p$ from month to month and latitude to latitude (c,d) is considerably larger (and comparable to the uncertainty of 0.015) and masks para-H$_2$ trends at 300 mbar. }
\label{temp_fpara}
\end{centering}
\end{figure}

More than 3000 $T(p)$ and $f_p(p)$ profiles were derived from the CIRS spectra, representing zonal-mean conditions for each latitude and month.  These are plotted in Fig. \ref{temp_fpara} to show the scatter of the measurements and the hemispheric asymmetries.  Numerous outliers exist that could significantly corrupt the interpretation of a particular date and latitude - this could be due to (i) unexpected longitudinal variability so that the spectrum does not accurately represent a zonal average; (ii) calibration baseline offsets that are sometimes produced when the conditions vary between target CIRS spectra and the calibration spectra; or (iii) the combination of spectra observed from different orbital distances, and hence different fields of view averaging over spatial gradients.  These spectra often appear to be realistic, and are interpreted literally by the retrieval algorithm to produce a good fit.  However, they produce retrieved temperatures and abundances that deviate noticeably from the general trends, and can be removed by filtering the results.  We filtered out results where (i) $T(p)$ and $f_p(p)$ profiles exhibited substantial oscillations with altitude; (ii) the tropopause $f_p$ exceeded 0.5; and (iii) the retrieved NH$_3$ and PH$_3$ scale factors deviated from the \textit{a priori} by more than $2\sigma$.  This removed $<5\%$ of the retrieved results.  

\begin{figure*}
\begin{centering}
\noindent\makebox[\textwidth]{\includegraphics[width=\paperwidth]{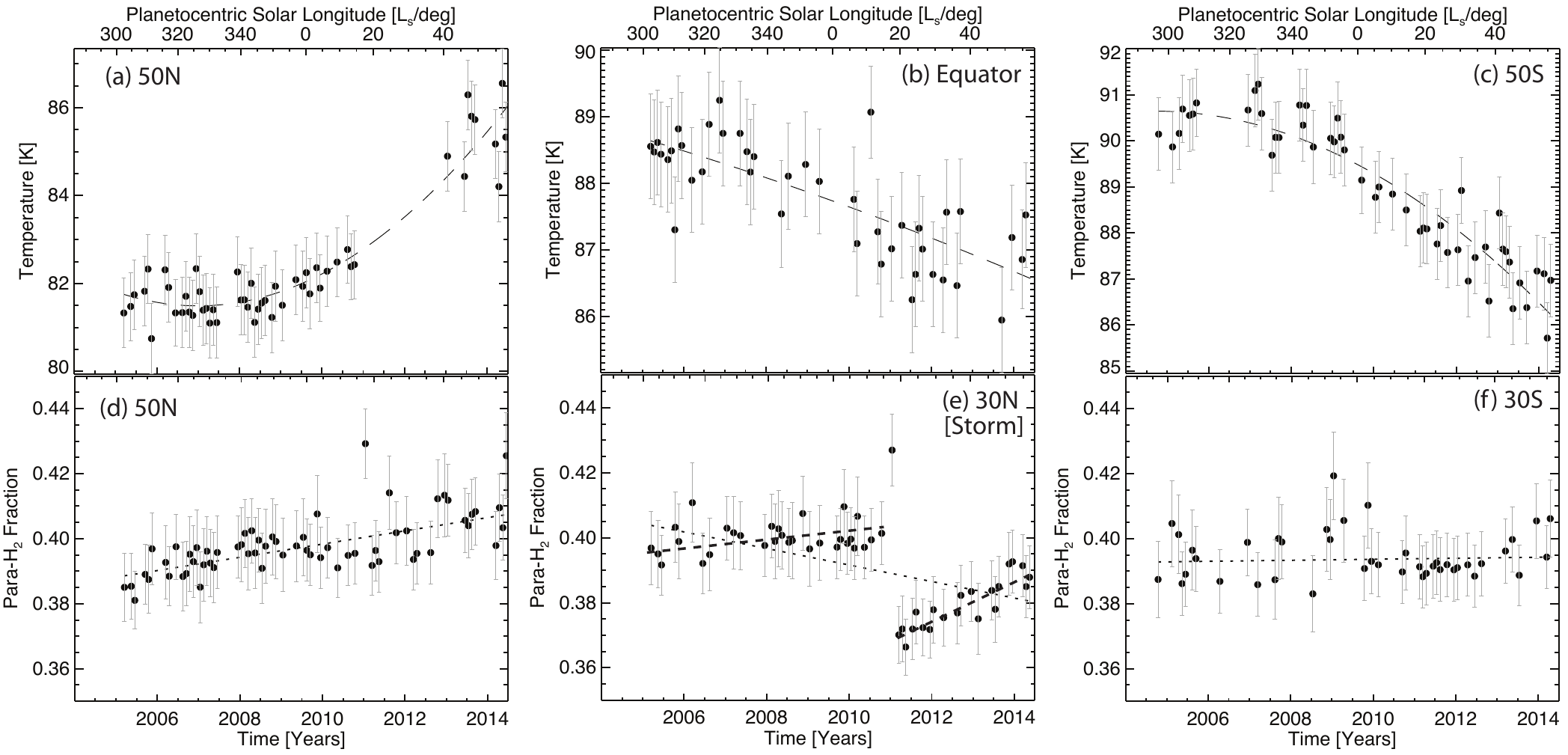}}
\caption{Examples of the polynomial fitting to temporal trends observed for temperature at 100 mbar (quadratic fits in panels a,b,c) and para-H$_2$ at 300 mbar (linear fits in panels d,e,f). In the top row, the comparison between $50^\circ$N and $50^\circ$S shows examples of seasonal warming and cooling, whereas the equator shows the effects of dynamics cooling the equatorial tropopause.  The para-H$_2$ examples show a case where $f_p$ is increasing with time (d, $50^\circ$N); staying stable (f, $30^\circ$S); and an example where the linear fit is inappropriate (e, $30^\circ$N) near the location of the 2010-11 storm eruption \citep{14achterberg}. In this latter case, dashed lines show a two-step regression line to show how it deviates from the 10-year trend.}
\label{temporal}
\end{centering}
\end{figure*}

For the remaining data, the temporal coverage of a particular latitude was often sparse and incomplete (Fig. \ref{coverage}), so we followed \citet{15fletcher_poles} in using smooth functions to interpolate over the irregularly-spaced data to `reconstruct' the temperature and para-H$_2$ variability.  We required a quadratic function to reproduce the 10-year variability of temperatures at each altitude, whereas a linear function was found to be sufficient for the compositional variability.  Examples of these temporal trends are shown for a selection of latitudes in Fig. \ref{temporal}.  Note that these smooth functions are for interpolation only, and are not intended for extrapolation outside of the time period covered by the CIRS data.  Clearly we have no knowledge of Saturn's true temperatures during these temporal and spatial gaps, but we assume that the temperature and para-H$_2$ continue to follow the identified trends.  These gaps are fully accounted for in the uncertainty on the interpolation functions and the reconstructed fields.

The interpolation functions were used to create `movies' of the zonal-mean $T(p)$ and $f_p(p)$ over the duration of the observations, available in our Supplementary Online Material.  Annually-averaged snapshots of the temperature are shown in Fig. \ref{temp_snapshot}; Fig. \ref{Tcontour} shows temperature trends as a function of latitude and time at 100 mbar and 400 mbar; Fig. \ref{curvature} shows the derivative of the lapse rate ($d^2T/dz^2$) for temperature profiles shown in Fig. \ref{Tknee}; and annual averages for the para-H$_2$ fraction and disequilibrium are shown in Fig. \ref{parah2_snapshot}.

Optimal estimation provides a formal estimate of the uncertainty on the retrieved profiles that depends on uncertainties in the measurements, \textit{a priori} and forward model, and also on the degeneracies between the retrieved parameters.  Examples of these individual uncertainties are shown in \mbox{Fig. \ref{temporal}}, and are used to weight the regression required to define the interpolation functions.    Propagation of error from the retrieved quantities to the derived quantities - the `reconstructed' temperature and para-H$_2$ distributions in \mbox{Section \ref{results}} and the winds, stratification, potential vorticity gradients and atmospheric refractivity in \mbox{Section \ref{discuss}} - is a considerable challenge.  Our approach is to use a Monte Carlo (MC) simulation to assess the potential range of these derived quantities.  Each of the 3000+ $T(p)$ and $f_p(p)$ profiles was perturbed within the retrieval uncertainty envelope, using a Gaussian perturbation with (i) a peak altitude selected at random from a uniform distribution of pressures between 100 and 800 mbar; and (ii) an amplitude calculated by multiplying the retrieval uncertainty by a number randomly selected from a normal distribution (i.e., the temperature profile remains within the $1\sigma$ retrieval uncertainty range 68\% of the time).  We then recalculated the regression lines for each time series to produce a new version of the reconstructed profiles, and repeat the derivation of the thermal winds, buoyancy frequency, potential vorticity gradients and index of refraction (see \mbox{Section \ref{discuss}}).  This process was repeated 100 times with random perturbations, and the resulting reconstructions are compared to provide a range of uncertainty in each parameter discussed in the following sections.  Note that this error analysis will not identify `unknown unknowns', such as calibration biases or inadequacies in our forward model.

\subsection{Tropospheric Temperature Changes}

\begin{figure}
\begin{centering}
%\noindent\makebox[\textwidth]{\includegraphics[width=\paperwidth]{figs/temp_snapshot.pdf}}
\includegraphics[angle=0,scale=0.7]{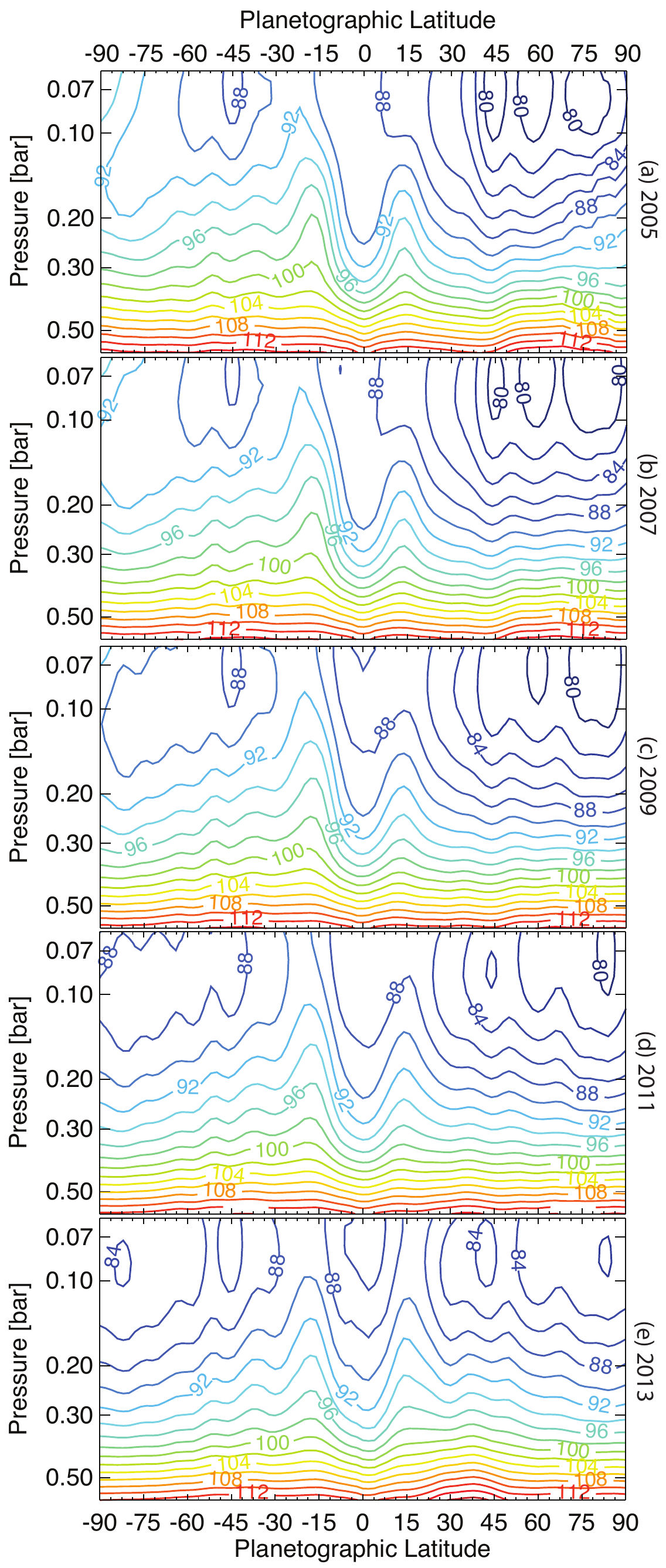}
\caption{Snapshots of the zonal mean temperatures reconstructed from the sparsely-gridded temperature retrievals.  Temperatures are averaged over a full year, and were chosen to represent Saturn's equinox (2009) $\pm4$ years.  The full time-lapse sequence is available in the supplemental online material.  }
\label{temp_snapshot}
\end{centering}
\end{figure}

\begin{figure}
\begin{centering}
%\noindent\makebox[\textwidth]{\includegraphics[width=\paperwidth]{figs/temp_snapshot.pdf}}
\includegraphics[angle=0,scale=0.7]{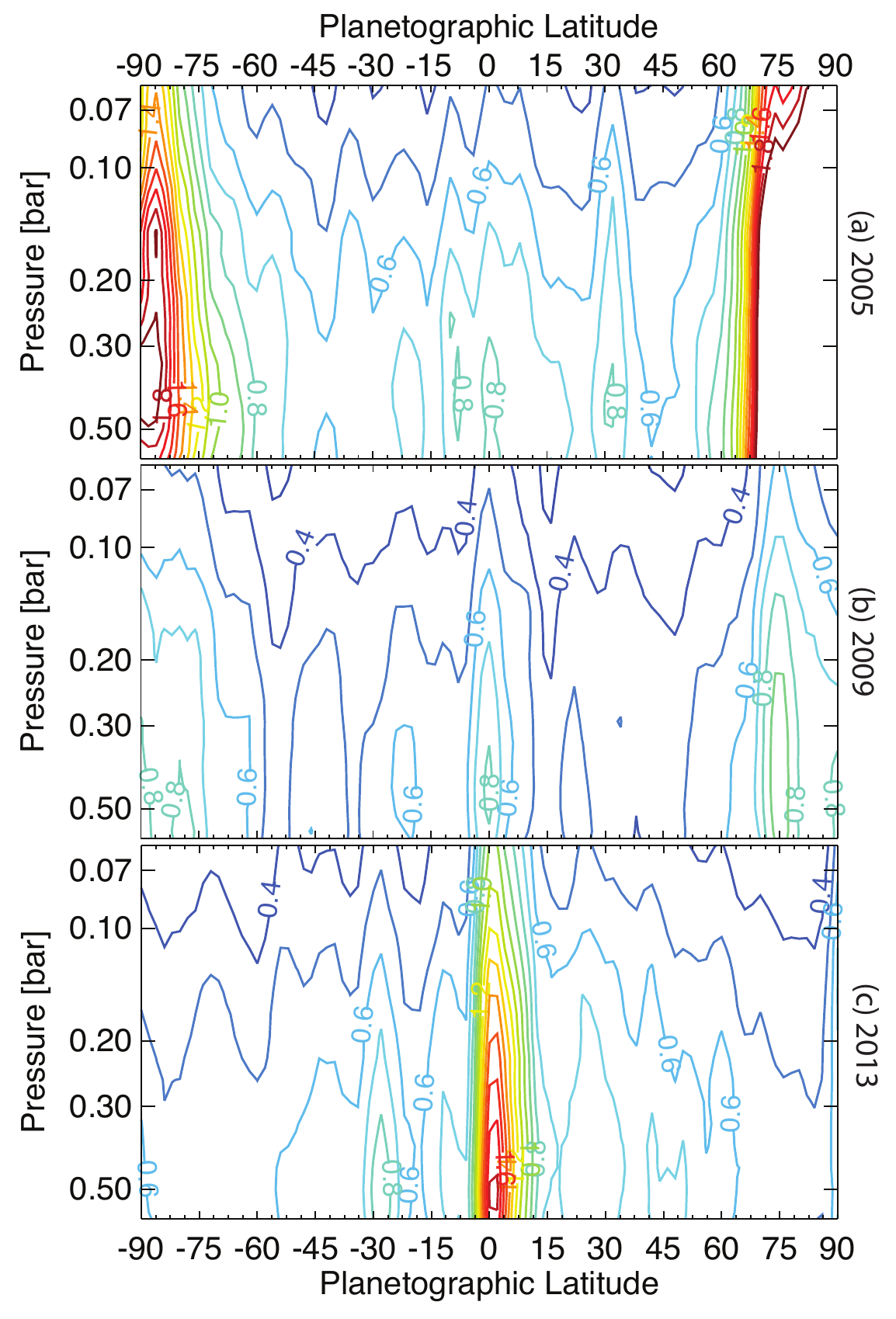}
\caption{Temperature profile uncertainties in Kelvin calculated via an MC simulation in 2005, 2009 and 2013 for comparison with the changes observed in Fig. \ref{temp_snapshot}. Given the nature of uncertainties in a regression line, the errors are time dependent, being smallest in the centre of the time series (2009). Contours are spaced every 0.1 K. }
\label{temp_error}
\end{centering}
\end{figure}

\begin{figure}
\begin{centering}
\includegraphics[angle=0,scale=0.7]{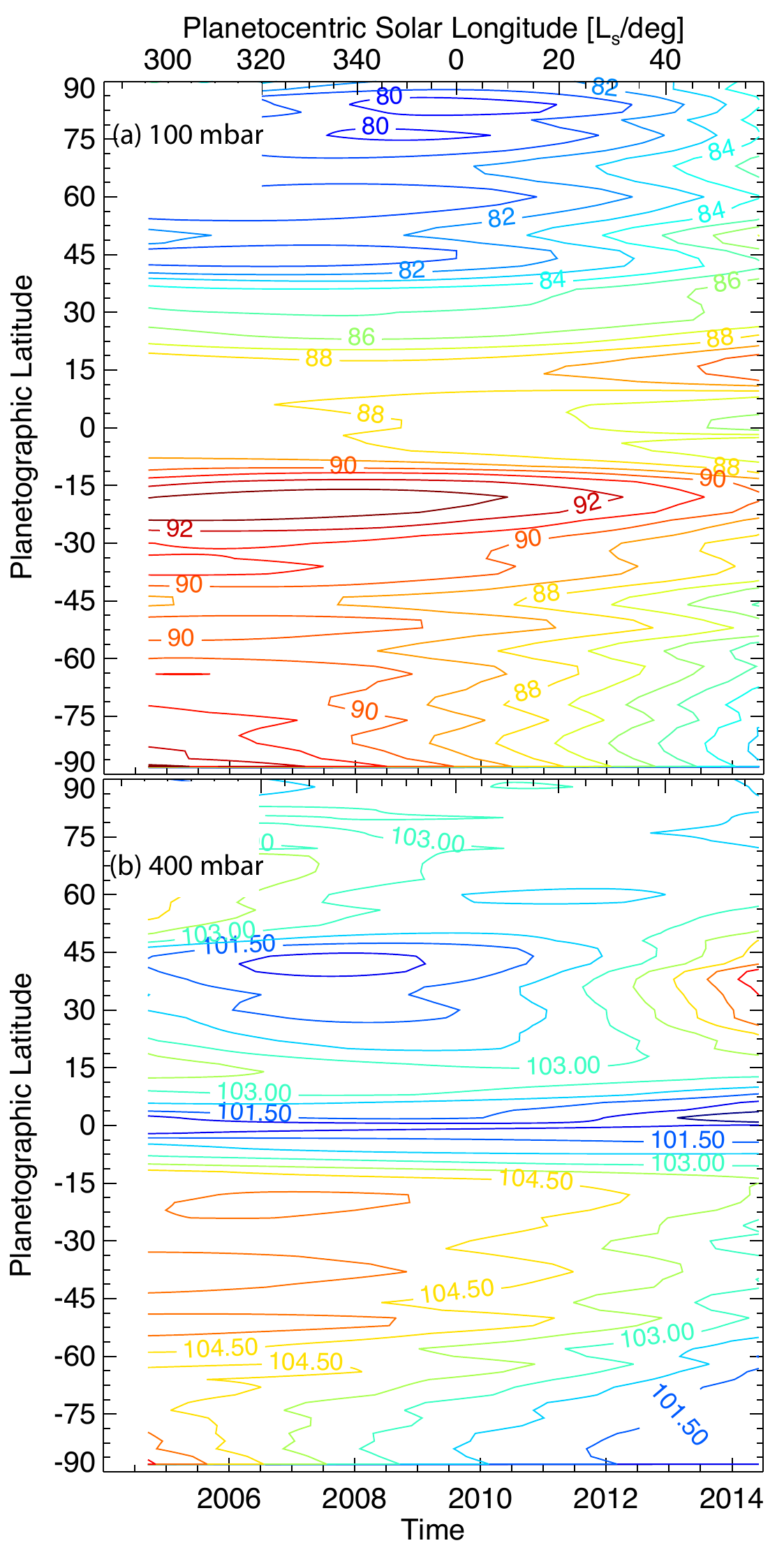}
\caption{Contour plot of temperature versus latitude and time for two tropospheric levels, 100 mbar and 400 mbar.  These figures were generated from the polynomial-smoothed `reconstructed' temperature fields, and are useful to show trends over the ten year span of the observations.  Contours are spaced every 1.0 K for 100 mbar and every 0.75 K for 400 mbar to aid visualisation.  Uncertainties on these temperatures are shown in Fig. \ref{temp_error}. }
\label{Tcontour}
\end{centering}
\end{figure}

% Changes in the southern hemisphere
Saturn's reconstructed tropospheric temperatures are shown in \mbox{Fig. \ref{temp_snapshot}}, , with the uncertainty estimated by the MC simulation shown in \mbox{Fig. \ref{temp_error}}.  Trends with time are shown in \mbox{Fig. \ref{Tcontour}}.  The nature of uncertainties in regression lines means that the errors are time dependent, being smallest near the centre of the time series (near equinox in 2009) and larger towards the ends (2005 and 2013).  Uncertainties are larger at higher pressures where the sensitivity of the spectrum to temperature changes is lower, and at high latitudes where the gaps in temporal coverage are larger (\mbox{Fig. \ref{coverage}}).  Indeed, uncertainties at northern high latitudes in 2005 are large ($>2$ K) because these regions weren't observed by CIRS until late 2006, and the interpolation functions are not meant to be used outside of these ranges.  Furthermore, as Cassini typically observed from Saturn's equatorial plane, the rings obscure the equator and result in gaps in the time series (and larger uncertainties) at low latitudes. Over most of the planet the uncertainties are in the 0.4-0.8K range, as expected from the individual retrievals.

Upon Cassini's arrival at Saturn, tropopause temperatures showed a notable $10\pm2$ K asymmetry between southern and northern high latitudes (Fig. \ref{temp_snapshot}a).  This asymmetry has been reversing throughout the duration of the mission, continuing the trend reported by \citet{10fletcher_seasons} using mid-infrared CIRS data.  Southern hemisphere temperatures have generally cooled with the onset of autumn, although there are suggestions for a maximum in tropopause temperatures ($92\pm2$ K) in the $20-40^\circ$S region in 2006, $\approx4$ years after solstice (Fig. \ref{Tcontour}).   In 2006, 100-mbar temperatures were less than 90 K only in the small region from $40-60^\circ$S, but by 2012 the entire southern hemisphere had cooled below 90 K, with temperatures less than 85 K poleward of $45^\circ$S by 2014.    

% Changes in the northern hemisphere
In the northern hemisphere, Fig. \ref{Tcontour} shows the 100-mbar temperatures continuing to cool through 2007 and 2008, but then begin to warm as spring approaches. The lowest temperatures at the tropopause ($78\pm1.5$ K near 80 mbar) were detected several years after the winter solstice (2002).  The precise timing of the temperature minimum varies as a function of altitude and latitude, with northernmost latitudes experiencing temperature minima near the spring equinox in 2009 \citep[$\approx7$ years after solstice, consistent with][]{15fletcher_poles}, whereas the minimum occurred in 2007 at 45$^\circ$N ($\approx5$ years after solstice, Fig. \ref{Tcontour}).  At higher pressures, the temperature minimum did not occur until early spring (a minimum north polar temperature of $87\pm1$ K at 220 mbar in 2010-2011 compared to $90\pm1$ K by 2014), consistent with the increasing atmospheric lag times with depth. Indeed, Fig. \ref{temp_fpara}b shows that 330-mbar temperatures poleward of $75^\circ$N were relatively constant over the whole observation period.  As of 2014 Fig. \ref{temp_fpara}a shows that the northern tropopause was cooler than 85 K for all latitudes poleward of $45^\circ$N (with coolest tropopause temperatures of $82\pm1$ K in the north polar zone near 80-85$^\circ$N), mirroring temperatures in the southern tropopause region poleward of $45^\circ$S.  This suggests that the summertime asymmetry has now vanished and the two hemispheres show a distribution of tropopause temperatures that is approximately symmetric about the equator.

% Equator
The equatorial region, broadly defined as all latitudes equatorward of $\pm20^\circ$, undergoes more unusual changes driven by dynamics.  At 220 mbar, the $5\pm1$ K contrast between the warm SEB and NEB observed at the start of the mission has now equalised (with 220-mbar temperatures of $95\pm0.5$ K, in contrast to equatorial temperatures of $89\pm0.7$ K, using uncertainties estimated by the MC technique).  However, at higher altitudes the equatorial tropopause has cooled from $89\pm0.5$ K to $86\pm0.5$ K over the ten year span of observations (Fig. \ref{Tcontour}).  The time-lapse movies suggest that this is due to the downward propagation of a cold temperature anomaly from the lower stratosphere, and such descent has been previously observed in the mid-stratosphere (0.01-20 mbar) by authors studying Saturn's semi-annual oscillation \mbox{\citep[SAO, e.g.,][]{11guerlet, 11schinder}.}  Note that the air itself is not descending, but the pattern of warm and cool anomalies that result from complex wave-mean-flow momentum interactions that vary with time and space as the windshear changes.  Indeed, the cool anomaly could be the result of local uplift and adiabatic expansion over a narrow altitude range.  The variation in the equatorial tropopause suggests that this oscillating zonal-mean wind and temperature structure is also impacting the 50-150 mbar region.  In the deeper atmosphere, the equatorial temperatures are stable with time.  This zone remains one of the coolest regions compared to other latitudes, consistent with the idea of uplift and adiabatic expansion at low latitudes, mirrored by subsidence and warming over the neighbouring SEB and NEB.

%The actual waves driving the SSAO (via wave-mean flow interactions where the waves break) would be upward propagating.

% Warming in the deep troposphere
Finally, two anomalously warm northern hemisphere regions ($T>114$K at 600 mbar) are visible in Fig. \ref{temp_snapshot} in the deeper troposphere.  The first region, the `bulge' observed at $50-70^\circ$N in 2005 that was identified in Cassini/CIRS inversions by \citet{07fletcher_temp}, cooled by 3-5 K over the ten year span of the data so that the latitudinal gradients in temperature ($dT/dy$) have vanished.  The disappearance of this winter anomaly will have implications for changing potential vorticity gradients in Section \ref{discuss}.  At the same time, a new region of $T>114$ K developed in the $25-45^\circ$N region from 2010 onwards.  This warming has been documented by \citet{14achterberg} and is associated with the eruption of Saturn's northern springtime storm.  However, a similar warm region was noted one Saturn year earlier by \citet{82hanel} and \citet{83conrath}, and we speculate that this deep atmospheric warming is a dynamical response to the changing season.    There appear to be no comparable regions in the southern hemisphere during this time period, with 600-mbar contours remaining relatively uniform with latitude.

\subsection{Aerosol Heating Effect}

\begin{figure*}
\begin{centering}
\noindent\makebox[\textwidth]{\includegraphics[width=\paperwidth]{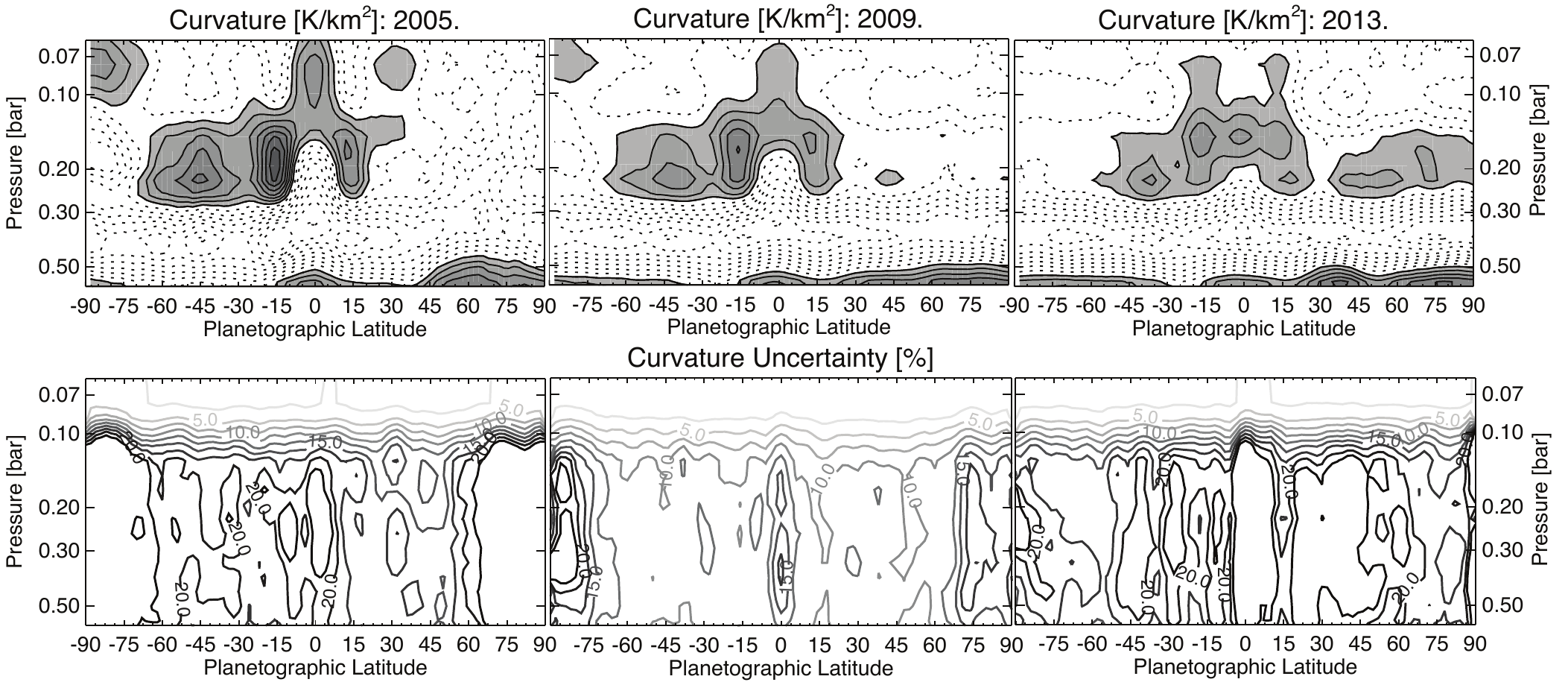}}
\caption{The curvature of the temperature profile as a function of time, used as a proxy for the size of the aerosol-related temperature perturbation in Saturn's upper troposphere.  Contour intervals are spaced by 0.002 K/km$^2$, with grey shading used to indicate regions of strong negative curvature (i.e., the strongest thermal perturbation).  Dotted contours indicate regions of positive curvature. Contours are shown as annual means for the equinox $\pm5$ years, with the full time-lapse sequence available in the supplementary materials.  Regions of negative curvature at $p>500$ mbar are spurious, caused by the retrieved $T(p)$ relaxing to the prior in a region of low spectral information content.  The lower panel shows the estimated uncertainty on the curvature, shown as a percentage of the 0.002 K/km$^2$ contour spacing (i.e., typical values of 10-25\% of a contour interval.}
\label{curvature}
\end{centering}
\end{figure*}

\begin{figure}
\begin{centering}
\includegraphics[angle=0,scale=0.7]{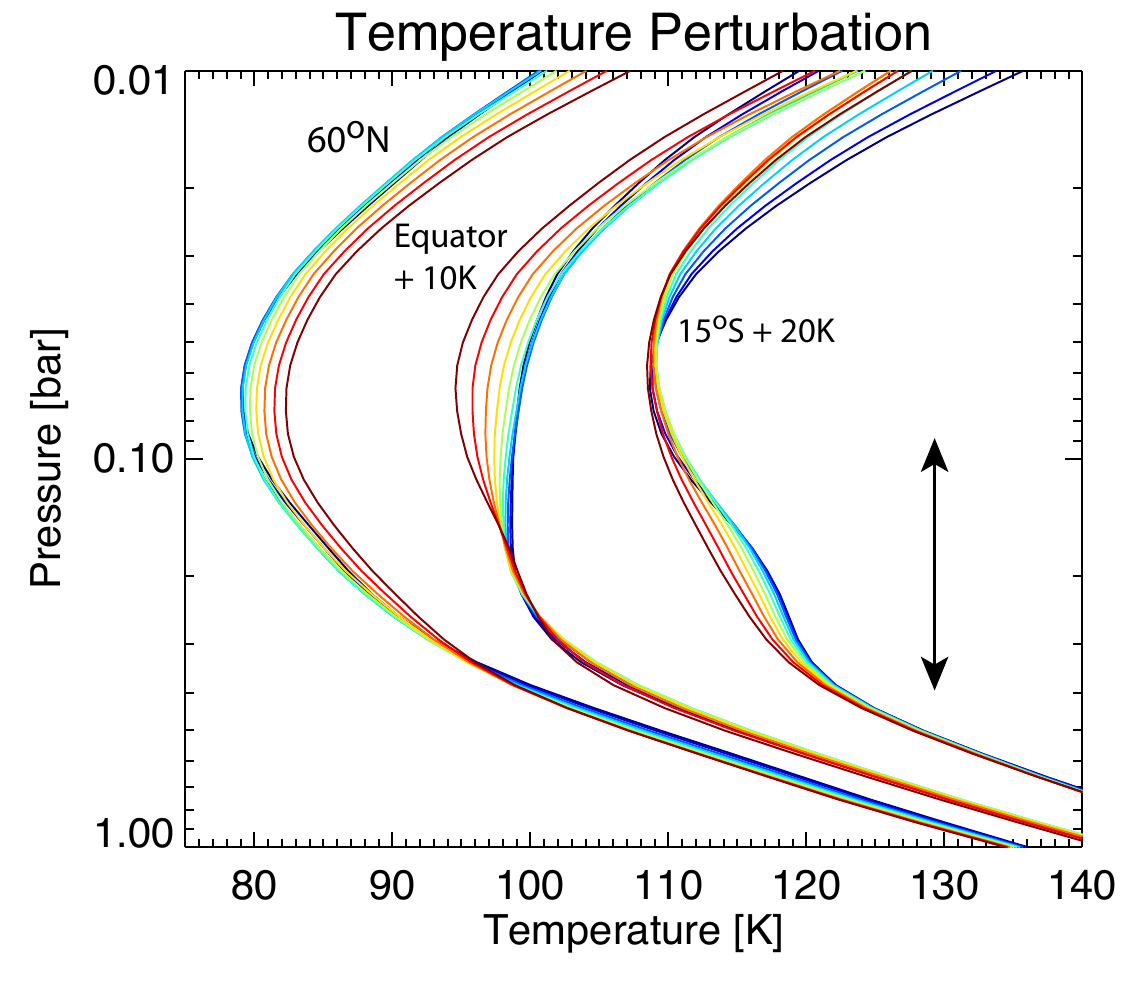}
\caption{Examples of the reconstructed $T(p)$ profiles showing the location of the `kink' related to aerosol heating in Saturn's upper troposphere.  One curve is shown for each of the ten years.  Three latitudes (the $60^\circ$N, the equator and $15^\circ$S) are shown, with 10 and 20-K offsets applied to the latter two.  Blue curves indicate profiles from early in the mission, red curves indicate curves from later in the mission, showing how the curvature of the $T(p)$ profiles has altered over time.}
\label{Tknee}
\end{centering}
\end{figure}

The region of localised heating in the 100-250 mbar region (Section \ref{intro}) can be quantitatively investigated via measurements of the `curvature' of the $T(p)$ profile by taking the second derivative, $d\Gamma/dz=d^2T/dz^2$ (where $\Gamma$ is the lapse rate).  This thermal perturbation is believed to be related to localised heating within Saturn's upper tropospheric haze \citep{07fletcher_temp,12friedson,14guerlet}, and showed a pronounced asymmetry early in the mission with a stronger perturbation in the southern summer hemisphere, and negligible in the north.  The magnitude of the curvature is shown in Fig. \ref{curvature}, and the reconstructed $T(p)$ profiles for three selected latitudes are provided in Fig. \ref{Tknee}.  Uncertainties tend to be magnified wherever derivatives are involved, so the MC simulation is used to show the uncertainty on this curvature in the bottom row of \mbox{Fig. \ref{curvature}}, in units of a single contour spacing (0.002 K/km$^2$).  The uncertainty in the curvature is typically less than 10\% of a contour interval, but can rise to 25\% and higher at the equator, poles and the storm perturbed latitudes due to gaps in the time series.  The peak negative curvature (grey shading) is located at the pressure of maximum thermal perturbation, and the peak positive curvature is located where the $T(p)$ profile tends to the deeper adiabat.  The inflection points (where the temperature profile changes the direction of curvature) are used to provide approximate upper and lower pressure limits on the thermal perturbation.  Note that regions of negative curvature near 600 mbar are likely to be spurious, related to relaxation of our retrieved $T(p)$ back to the \textit{a priori} profile.  Finally, the size of the curvature is sensitive to our choice of vertical smoothing in the original inversions \citep[which is kept the same for every retrieval, see][for details]{07fletcher_temp}, so we restrict our discussion below to qualitative trends in the curvature over time.

During southern summer, the largest thermal perturbation is observed over the SEB, and the feature is observed over much of the southern hemisphere and the NEB in the 150-250 mbar range.  The perturbation appears higher (70-150 mbar) and weaker over the equator, as observed by \citet{07fletcher_temp}, although uncertainties are highest at low latitudes.  There were no regions of negative curvature in the northern winter hemisphere poleward of $20^\circ$N.  The magnitude of the perturbation shrank but the location remained unchanged through to the spring equinox in 2009.  From 2008 onwards, regions of negative curvature began to appear in the northern hemisphere, starting in the cool zone near $45^\circ$N.  By 2010-11, the thermal perturbation was being detected throughout the northern hemisphere poleward of $35-40^\circ$N.  At the same time, the curvature had shrunk so much in the south that it vanished entirely from latitudes poleward of $45-50^\circ$S, giving the impression that the asymmetry observed at the start of Cassini's mission had reversed.  By 2014, the NEB and SEB exhibited thermal perturbations of the same relative size, and the `kink' was more prevalent in the north than in the south - a change larger than the uncertainties shown in Fig. \ref{curvature}.  Interestingly, the high extension over the equatorial zone vanished over the ten-year span of the data (the peak negative curvature was pushed deeper from near 80 mbar to near 150 mbar), which could be related to the cooling equatorial tropopause in response to Saturn's SAO wave train.  

% Relation to aerosol properties
With CIRS data alone, it is impossible to assess whether the growth of the thermal perturbation in the north is due simply to enhanced sunlight on an existing, static tropospheric haze \citep[as suggested by the circulation and transport model of][]{12friedson}, or whether the dominant effect is due to changes in the haze itself as spring progresses \citep[e.g.][]{05karkoschka}.  The latter is supported by colour changes observed by Cassini's cameras \citep{12edgington}, and a stark asymmetry in tropospheric haze opacity in 2006 was identified by Cassini's near-infrared spectrometer \citep[VIMS,][]{06baines_dps}, with negligible haze opacity in the hemisphere emerging from winter.  Growth of tropospheric aerosols in springtime was suggested by \citet{05karkoschka} from their analysis of Hubble data, and \citet{14guerlet} show that the size of the thermal perturbation would increase with tropospheric particle size and/or increased opacity.  However, a quantitative time series of Saturn's haze variability is required to fully understand the shifting location and size of this thermal perturbation.

\subsection{Para-Hydrogen}

\begin{figure*}
\begin{centering}

\includegraphics[angle=0,scale=0.8]{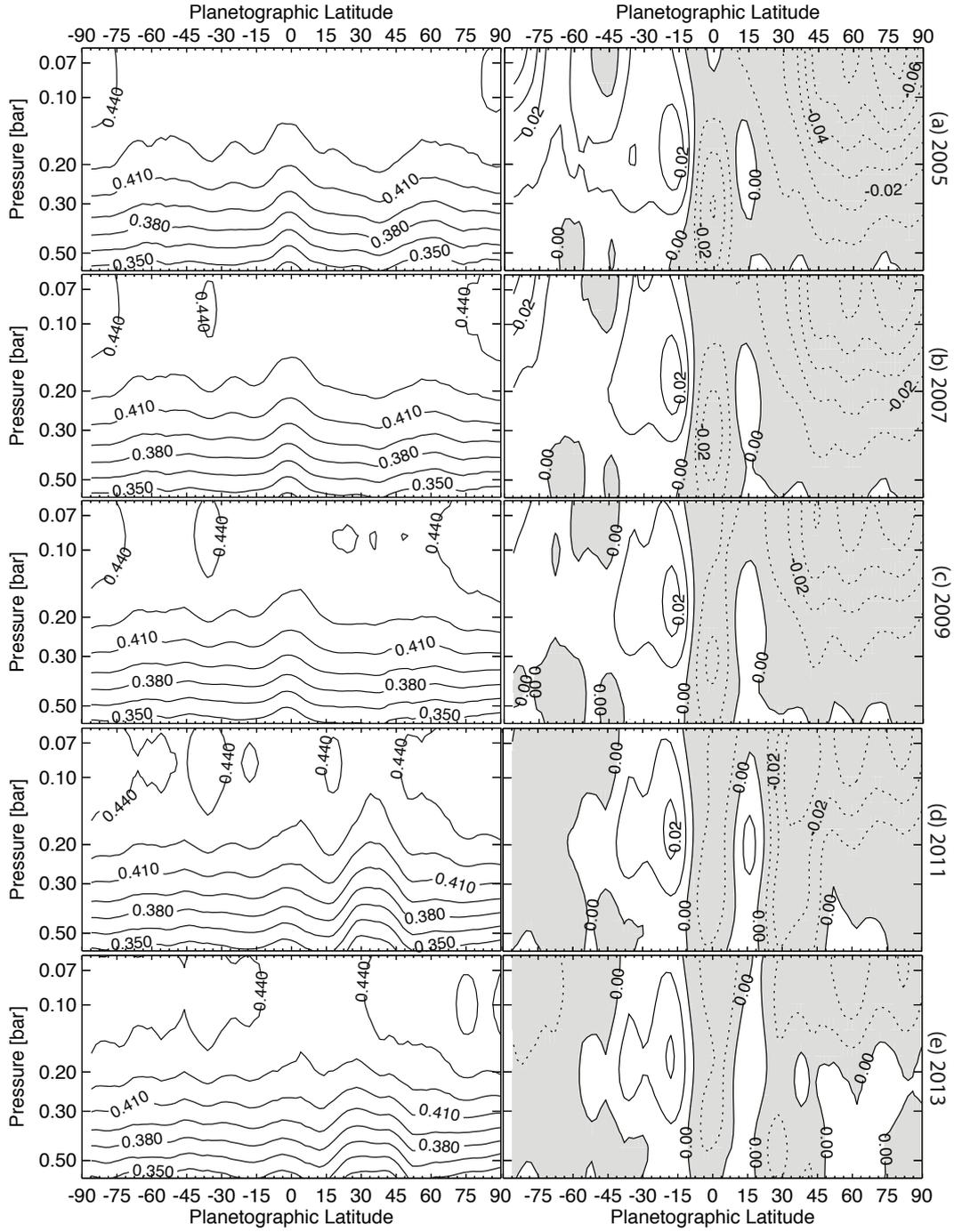}

\caption{Reconstructed para-H$_2$ fraction (left column) and degree of disequilibrium (right column) for five years:  Saturn's northern spring equinox (2009) and four years either side of it.  Uncertainties calculated via our MC simulation are shown in Fig. \ref{fperror}.  Contours of $f_p$ are spaced every 0.015,  equivalent to the largest uncertainty.  Contours of disequilibrium are spaced every 0.01, with sub-equilibrium conditions ($f_p<f_{eqm}$) indicated by the grey shading and dotted contours.  The full time-lapse sequence is available in the supplemental online material. }
\label{parah2_snapshot}
\end{centering}
\end{figure*}

\begin{figure}
\begin{centering}
\includegraphics[angle=0,scale=0.8]{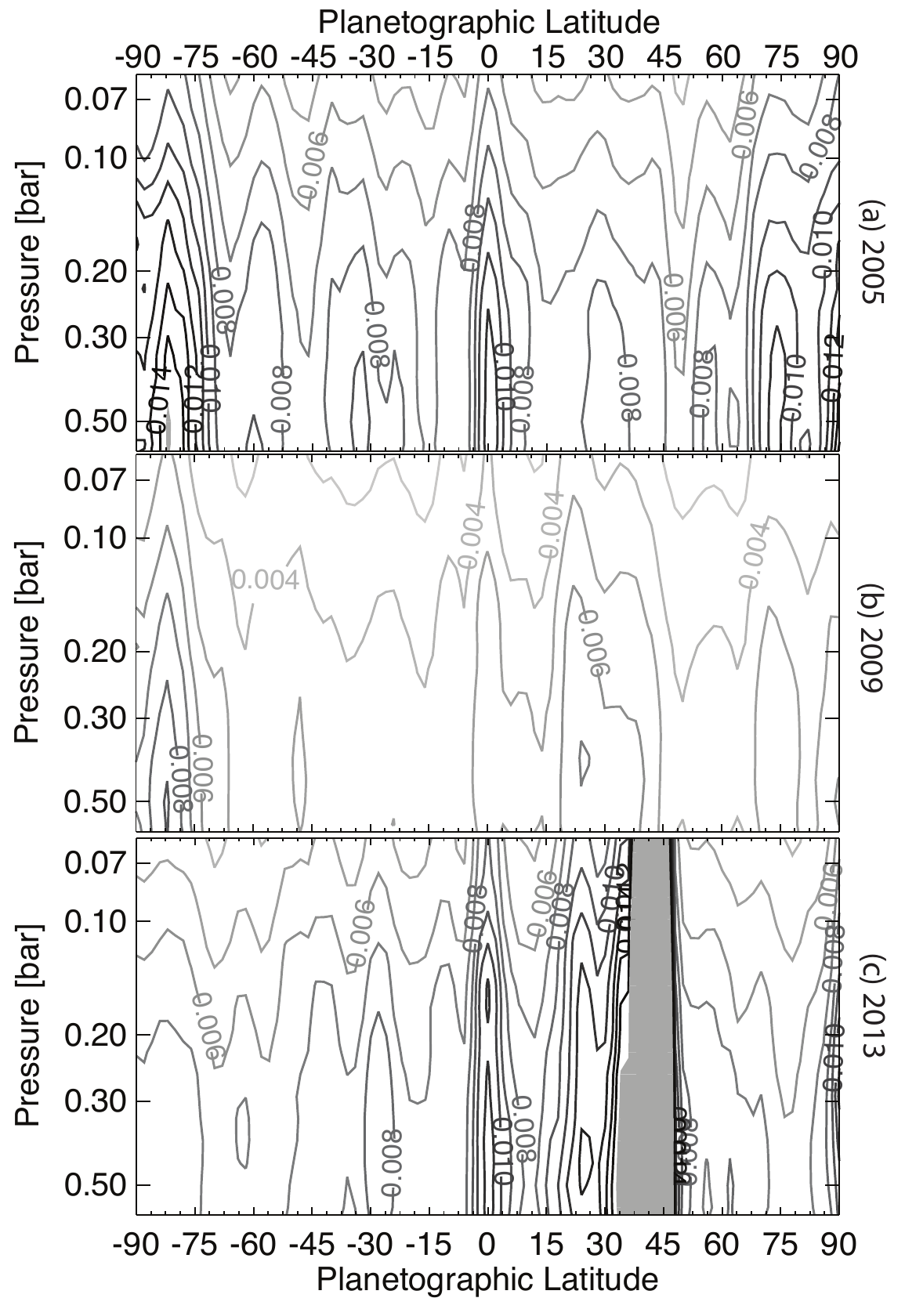}
\caption{Uncertainties on the reconstructed para-hydrogen fraction calculated from an MC simulation within the formal retrieval envelope.  The uncertainty is time dependent, and annually-averaged examples are shown for 2005, 2009 and 2013.  Contours are spaced every 0.001, with grey shading used to indicate where uncertainties exceed 0.015 (equivalent to the contour spacing in Fig. \ref{parah2_snapshot}), particularly following the 2010-2011 storm.  Uncertainties on the $f_p$ disequilibrium show the same spatial structure and similar amplitudes.}
\label{fperror}
\end{centering}
\end{figure}

\begin{figure}
\begin{centering}
\includegraphics[angle=0,scale=0.65]{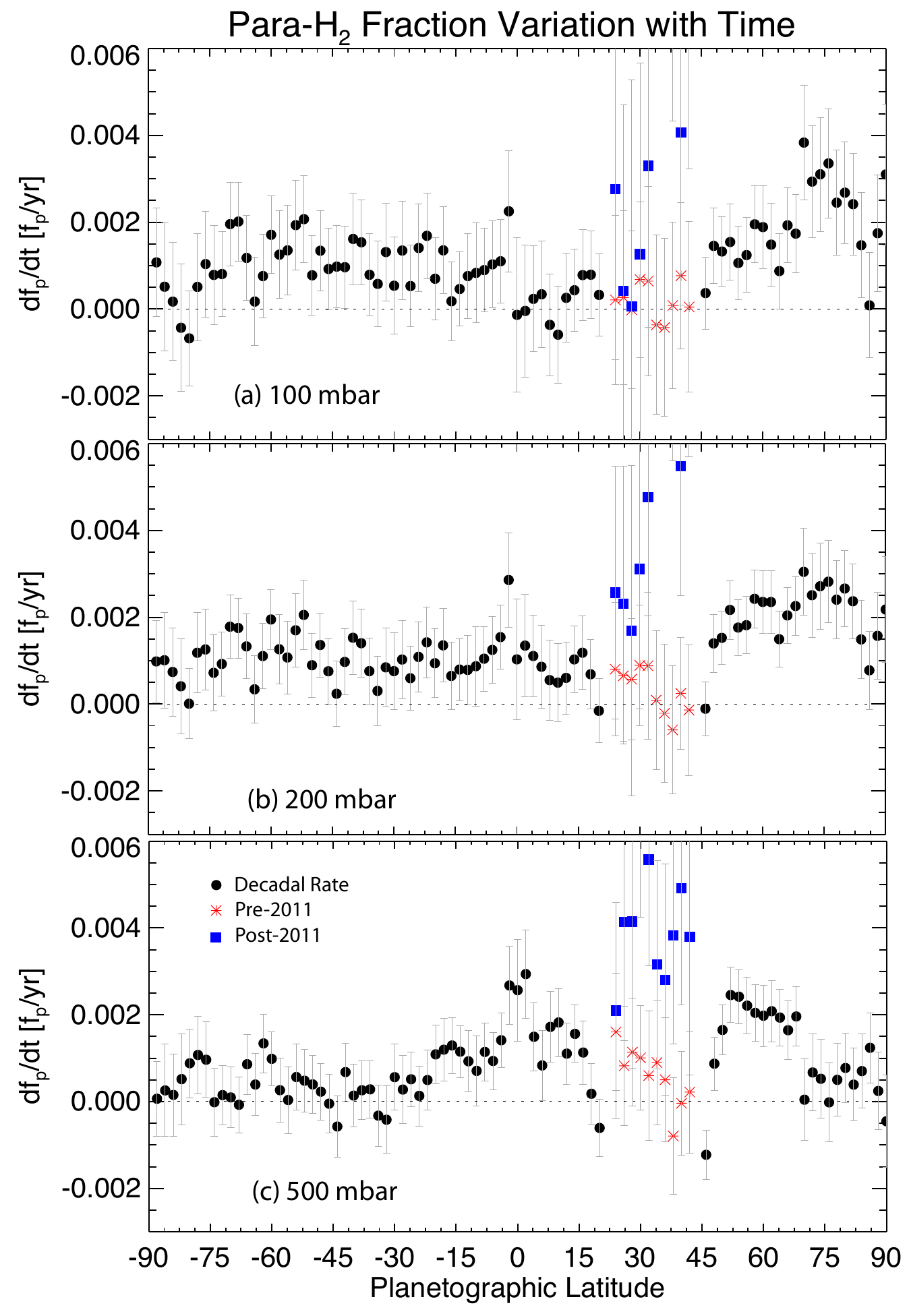}
\caption{Rates of change of para-H$_2$ fraction at three different altitudes:  (a) 100 mbar; (b) 200 mbar; (c) 500 mbar.  Black circles use a linear regression over the whole span of the data, red crosses between $15-45^\circ$N use only pre-2011 data for the regression to eliminate the springtime storm effects; blue squares use only post-2011 data to show the increased rates in the aftermath of the storm.  The rising $f_p$ at high northern latitudes is more apparent at lower pressures.  The horizontal dotted line shows where $f_p$ has been invariant with time.}
\label{dqdt}
\end{centering}
\end{figure}

Zonal-mean cross-sections for para-H$_2$ as a function of time were generated from the linear interpolation function, and five annual snapshots of the time-lapse movies are shown in Fig. \ref{parah2_snapshot}.  As we found for the temperature field, uncertainties on $f_p$ estimated by the MC simulation (\mbox{Fig. \ref{fperror}}) are time-dependent, being smallest near the middle of the time series ($\approx0.005$ for 300 mbar in 2009) and growing at either end.  Fig. \ref{temporal} indicated that a single linear rate was not appropriate at the storm latitudes in 2011-2013 ($15-45^\circ$N), where $f_p$ exhibited a sharp drop \citep{14achterberg} (see Fig. \ref{parah2_snapshot}(d)), so we use a 2-step linear function (pre-2011 and post-2011) for northern mid-latitudes.   These linear rates of change of para-H$_2$ ($d(f_p)/dt$) and their uncertainties are shown in Fig. \ref{dqdt}, broken down into three altitude ranges.   The smaller number of points contributing to the regression lines post-2011 result in larger uncertainties ($>0.015$) in the reconstructed $f_p$ at the storm latitudes in Fig. \ref{fperror}c.  Furthermore, we caution that CIRS is only sensitive to para-H$_2$ in the 100-550 mbar range \citep{07fletcher_temp}, and cannot constrain $f_p$ at altitudes above the tropopause.  Finally, we remind the reader that the quantitative values of $f_p$ and $f_{eqm}$ are sensitive to Saturn's helium abundance, for which we use the value of \citet{00conrath}.  For these reasons we restrict our discussion to changes to the para-H$_2$ fraction with time, rather than the absolute values.

Fig. \ref{dqdt} shows that the majority of the retrievals at 100 mbar are consistent with no change to $f_p$ over the ten-year data span for much of the mission (uncertainties on $d(f_p)/dt$ mean that they are consistent with zero).  There is a hint that $f_p$ has increased over the southern hemisphere by 0.01-0.02, but this is at the level of the uncertainty described above.  The only significant change at 100 mbar is in the $f_p$ poleward of $45^\circ$N, where para-H$_2$ has increased by $0.03\pm0.01$ over the course of the mission.  At intermediate pressures (200 mbar) near the peak of the contribution function, the southern hemisphere increase of 0.01 is still apparent, as is the 0.02-0.03 increase poleward of $45^\circ$N.  Before 2011, $f_p$ had shown an increase with time near $30^\circ$N (Fig. \ref{temporal}e and the red crosses in Fig. \ref{dqdt}), but $f_p$ decreased abruptly between $15-45^\circ$N at the start of 2011 in association with upwelling during the springtime storm \citep{14achterberg}.  In the aftermath of the storm, $f_p$ continued to rise but at a faster rate than before (blue squares in Fig. \ref{dqdt}).  At higher pressures (500 mbar) there is no change over the southern hemisphere, a 0.01-0.02 increase in $f_p$ at the equator, a general 0.02-0.03 increase over northern latitudes (particularly $50-70^\circ$N) punctuated by effects of the northern storm, and negligible change over the north polar region.  

Fig. \ref{parah2_snapshot} and \ref{dqdt} reveal that Saturn's tropospheric para-H$_2$ fraction has generally increased over the northern hemisphere, but that this trend is modified in both the storm region and the northern high-latitudes.   The latter can be interpreted as high-latitude tropospheric subsidence in the spring season, consistent with similar subsidence identified in the stratosphere by \citet{15fletcher_poles}.  At higher altitudes, \citet{15fletcher_poles} used the rising north polar temperatures and enhancements in hydrocarbons ethane and acetylene to infer downward motion over the 75-90$^\circ$N region.  This subsidence may persist into the upper troposphere for pressures less than 300-400 mbar, but not as deep as 500 mbar where $f_p$ was not observed to be changing.  Furthermore, analysis of the Voyager/IRIS data by \citet{98conrath} also suggested a higher $f_p$ in the northern hemisphere just after spring equinox, and suggested that this was due to a downward displacement.  They found large values of $f_p$ centred on $60^\circ$N (confirmed by our reanalysis in Section \ref{iris}), exactly where Fig. \ref{dqdt}c shows some of the largest temporal trends at 500 mbar outside of the storm latitudes.  The increase we see at $60^\circ$N has been continuous over Cassini's observations, unperturbed by the effects of the storm.  The changes in this region, and in the north polar region, are likely to be due to responses of the $f_p$ distribution to dynamics.  Conversely, the small 0.01-0.02 rise in $f_p$ at the tropopause over the southern hemisphere as autumn approaches could be a seasonal response related to shorter equilibration times in the hazier southern hemisphere.

The changing para-H$_2$ disequilibrium observed in Fig. \ref{parah2_snapshot} reflects primarily the changing temperature structure (and therefore distribution of $f_{p,eqm}$) rather than actual changes in $f_p$ induced by chemical conversion or transport.   Uncertainties on the disequilibrium are slightly larger than those shown in \mbox{Fig. \ref{fperror}} for $f_p$ alone, as they must account for both temperature and $f_p$ uncertainties.  Nevertheless, they are generally smaller than 0.01 (the contour spacing of the disequilibrium plots in \mbox{Fig. \ref{parah2_snapshot}}.  At the start of the mission, \citet{07fletcher_temp} identified super-equilbrium conditions ($f_p>f_{p,eqm}$) over the southern summer hemisphere and sub-equilibrium conditions ($f_p<f_{p,eqm}$ at the equator and over the northern winter hemisphere.  They speculated that this difference was not due to large-scale inter-hemispheric circulation (i.e., upwelling in the north and subsidence in the south), but rather due to the enhanced efficiency of conversion of para-H$_2$ to ortho-H$_2$ in the summer hemisphere where there was a greater density of aerosols to serve as catalysts \citep[e.g.,][]{03fouchet}.  Studying the snapshots of the disequilibrium in Fig. \ref{parah2_snapshot}, we see that neither of these interpretations is correct.  The super-equilibrium conditions have disappeared in the south as the temperatures dropped, tending towards equilibrium and sub-equilibrium (poleward of $45^\circ$S in 2014) as $f_{p,eqm}$ rose to match the value of $f_p$ that was detected in southern summer.  Likewise, in the north the sub-equilibrium conditions have declined as the temperature rose, so that equilibrium and super-equilibrium conditions are detectable between $45-75^\circ$N for $p>200$ mbar.  In this case, $f_{p,eqm}$ decreased with the warming springtime temperatures to match the $f_p$ found there during northern winter. In short, the large-scale asymmetries in the disequilibrium are simply driven by the temperature-dependent changes to $f_{p,eqm}$, and not due to changes in $f_p$ itself.  Neither large-scale circulation nor aerosol catalysis are required to explain the asymmetry in $f_p-f_{p,eqm}$.  

At low pressures, the detection of disequilibrium at certain points in Saturn's seasonal cycle is consistent with the findings of \mbox{\citet{98conrath}}, whereby the para-H$_2$ equilibration timescale greatly exceeds the timescales for thermal changes (radiative or dynamical), so the two spin isomers are `frozen in' and $f_p$ (but not $f_p-f_{p,eqm}$) traces the atmospheric circulation.  However, $f_p$ generally remains close to equilibrium at higher pressures, and we cannot rule out the possibility that surface catalysis on Saturn's aerosols plays a role in rapid interconversion of the two spin isomers \mbox{\citep[e.g.,][]{82massie}}.  As the timescales for equilibration (frozen or fast) influence the lapse rate, this degeneracy will influence the quantities derived in Section \mbox{\ref{discuss}}.     

The pattern of shifting disequilibrium is only broken where regional dynamics become important at the equator, poles and storm-perturbed latitudes.  The sub-equilibrium conditions at the equator and corresponding super-equilibrium conditions near $\pm17^\circ$ have persisted throughout the dataset, although there are signs that the local minimum in $f_p$ has weakened at higher pressures (e.g., Fig. \ref{dqdt}c).  This pattern was interpreted at the start of Cassini's mission as due to upwelling in the equatorial zone and subsidence in the neighbouring equatorial belts \citep[e.g.,][]{07fletcher_temp}.  The balance between upwelling and chemical equilibration could now be shifting, possibly due to weakened (or reversing) vertical motions influenced by the oscillating temperatures and winds at lower pressures.  At other locations the storm eruption caused $f_p$ to fall between $15-45^\circ$N to generate a corresponding sub-equilibrium region in Fig. \ref{parah2_snapshot}(d).  And at high northern latitudes, where the largest sub-equilibrium was revealed a the start of the Cassini mission, the high-latitude subsidence is causing the most rapid return to equilibrium conditions anywhere on the planet (a change in $f_p$ of 0.04-0.05).  

%%%%%%%%%%%%%%%%%%%%%%%%%%%%%%%%%%%%%%%%%%%%%%
%%%%%%%%%%%%%%%%%%%%%%%%%%%%%%%%%%%%%%%%%%%%%%
%%%%%%%%%%%%%%%%%%%%%%%%%%%%%%%%%%%%%%%%%%%%%%
\section{Voyager/IRIS Reanalysis}
\label{iris}

The reconstructed Cassini $T(p)$ and $f_p(p)$ span the same points in the seasonal cycle ($L_s=8.6^\circ$, April 2010, and $L_s=18.2^\circ$, February 2011) as the Voyager 1 and 2 flybys, respectively, and therefore permit a quantitative comparison of tropospheric conditions to search for evidence of inter-annual variability.  This has previously been investigated by \citet{13li} and \citet{14sinclair}, but using Cassini and Voyager datasets which were not obtained at exactly the same point in Saturn's season.  The Voyagers carried the Infrared Radiometer, Interferometer and Spectrometer (IRIS) instruments, with a Michelson interferometer providing 180-2500 cm$^{-1}$ spectra of the giant planets at a spectral resolution of 4.3 cm$^{-1}$ \citep{80hanel}.  We reanalyse the Voyager/IRIS Saturn data \citep{98conrath} to ensure that the same spectral model (in particular, the updated H$_2$-H$_2$ collision induced absorption), \textit{a priori} atmosphere and retrieval assumptions are used for both Voyager and Cassini datasets.   IRIS spectra were extracted from the expanded volumes available on NASA's Planetary Data System\footnote{http://pds-rings.seti.org/voyager/iris/expanded\_volumes.html}, and zonal-mean spectra were constructed from observations within $\pm2$ days of closest approach (November 13, 1980 and August 26, 1981 for Voyager 1 and 2, respectively).  Unlike Cassini's global mapping sequences, the Voyager data were acquired in individual latitude scans resulting in irregular latitudinal coverage and low spatial resolutions at mid-to-high latitudes.  Zonal-mean spectra were coadded in latitudinal bins $8^\circ$ wide and stepped every $2^\circ$, ensuring that the IRIS field of view was entirely on Saturn's disc.  Spectra were pre-filtered to remove any with a high degree of oscillations, and uncertainties were derived from the Voyager noise-equivalent spectral radiance (NESR), as described by \citet{14sinclair}.  In total, we used 990 Voyager-1 spectra (covering all latitudes) out of approximately 2100 spectra taken within $\pm2$ days of closest approach. Voyager-2 spectra had a limited latitudinal coverage (covering the northern hemisphere, with limited observations in the south) and proved more difficult to fit, possibly due to a misalignment of the interferometer \citep{82hanel,14sinclair}.  Furthermore, their Cassini-era counterpart coincides with a time where Saturn's northern latitudes were strongly perturbed by the 2010-2011 storm, so they were omitted from the remainder of this study.

We derive $T(p)$ and $f_p(p)$ from the 220-550 cm$^{-1}$ region of the IRIS spectra (encompassing the S(0) line and the long-wavelength edge of S(1)), supplemented by the 1230-1370 cm$^{-1}$ region to provide constraints on stratospheric temperatures \citep{14sinclair}.  The improved signal-to-noise ratio of IRIS near the S(1) line compared to CIRS, but the absence of data in the translational part of the spectrum below 220 cm$^{-1}$ (e.g., Fig. \ref{spxvar}), implies that IRIS offers different spectral constraints to CIRS for para-H$_2$, so caution must be taken in quantitative comparisons.  Even if both instruments were restricted to an identical spectral range, the different noise characteristics of IRIS versus CIRS would inevitably provide different measurement uncertainties in the retrieval process.

\begin{figure*}
\begin{centering}
\includegraphics[angle=0,scale=0.8]{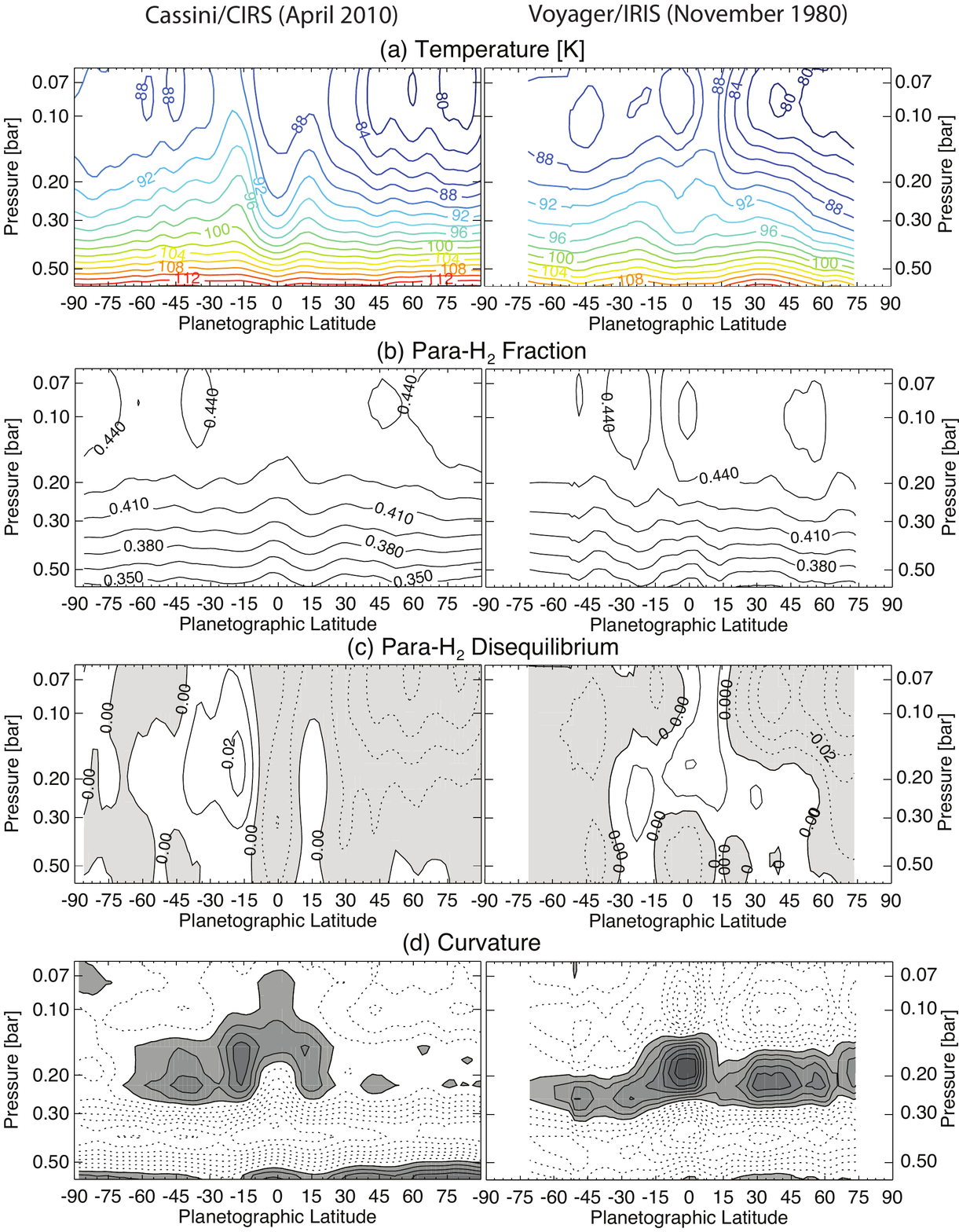}

\caption{Comparison of results from Voyager/IRIS (November 1980) with those from Cassini/CIRS (April 2010), interpolated to the same point in Saturn's seasonal cycle ($L_s=8.6^\circ$).  Panel (a) shows tropospheric temperatures with a 2-K contour spacing; panel (b) shows the para-H$_2$ fraction with a 0.015 contour spacing; panel (c) shows the deviation of $f_p$ from equilibrium in steps of 0.01; and panel (d) shows the curvature of the temperature profile in steps of 0.002 K/km$^2$.  Despite the similar season, stark differences are evident in the tropical regions and at mid northern latitudes, although these are close to the conservative uncertainty estimates (1 K on $T(p)$ and 0.02 on $f_p(p)$) for the two datasets.}
\label{iris_compare}
\end{centering}
\end{figure*}

From inversions of the Voyager-1 spectra, we confirm that Saturn's 1-mbar equatorial temperatures were colder than the neighbouring belts at this time period, supporting the conclusion of \citet{14sinclair} that Voyager and Cassini captured a different phase of Saturn's stratospheric oscillation (known as the semi-annual oscillation, SAO).  Tropospheric temperatures, the thermal perturbation due to aerosol heating, para-H$_2$ and the degree of disequilibrium are shown side by side for Voyager-1 and Cassini in Fig. \ref{iris_compare}, with the Cassini results interpolated to April 2010 to match $L_s=8.6^\circ$.  Typical errors on the IRIS $T(p)$ are 0.6-0.7 K between 100 and 600 mbar, 1.6-1.8 K at 1 mbar.  Typical errors on the $f_p(p)$ are 0.015-0.019 between 100-500 mbar. These are comparable to the uncertainties on the CIRS reconstructions in Fig. \ref{temp_snapshot} and Fig. \ref{parah2_snapshot}.

Tropopause temperatures in Fig. \ref{iris_compare}a show a similar asymmetry for both datasets between the cool north and warmer southern hemisphere, although IRIS found cooler (85-87 K) southern hemisphere temperatures than CIRS (88-90 K), whereas the northern tropopause temperatures (80-85 K) were roughly comparable between both epochs.  At pressures of 500 mbar and deeper, the IRIS inversions also produce systematically cooler temperatures by 3-4 K compared to the CIRS inversions.  Random uncertainties from the IRIS inversion and CIRS interpolation procedure are smaller than 1 K at these altitudes, so cannot explain this offset.  This could be a real change over the Saturnian year, but systematic calibration offsets between the two instruments have not been adequately assessed, and these would significantly hamper any attempt at quantitative comparison.  We therefore restrict our comparison to relative spatial contrasts.

The largest differences between the two epochs occur within $15^\circ$ the equator:  in the Voyager-1 observations the NEB appears warmer than the SEB by 2-3 K at 100 mbar, whereas the Cassini observations show the NEB to be cooler than the SEB by 1-2 K, qualitatively consistent with the results from \citet{13li}.  This difference is only apparent for $p<200$ mbar, but serves to alter the latitudinal thermal gradient around the equatorial zone and therefore the vertical shear on the zonal wind \citep{13li}.  The temperature comparison therefore suggests non-seasonal effects at Saturn's equatorial tropopause that disrupts the seasonal cycle.  Also notable is the $20-45^\circ$N warm region in the IRIS retrieval at 600 mbar \citep[also evident in Fig. 6 of][]{98conrath} that is absent from the CIRS retrieval (this was 8 months prior to the eruption of the springtime storm in December 2010).

Cassini's results showed that the thermal perturbation between 150-250 mbar featured a maximum curvature elevated over the equator, with the size of the perturbation decreasing in the southern autumn hemisphere and growing in the northern spring hemisphere.  However, the perturbation appeared larger in the north during the Voyager-1/IRIS observations than during the Cassini/CIRS observations, suggesting that the mechanism responsible for the heating may not be exactly reproducible from year to year.  Unfortunately, detailed comparisons of the aerosol loading between the Voyager and Cassini springtime epochs are not yet available to see if this ties to the thickness and coloration of Saturn's tropospheric hazes.  However, we caution that quantitative comparisons of the $T(p)$ curvature between the two datasets are difficult, as the lower-resolution IRIS data provides a different constraint on the spectral inversions compared to the higher-resolution CIRS data.

Finally, the para-H$_2$ fraction shows an asymmetry between the northern and southern hemispheres in the Voyager epoch, whereas the Cassini results are more symmetric with a generally higher value of $f_p$ in the northern spring hemisphere (by 0.02-0.03).  There are, however, several quantitative differences between the two epochs:  Voyager observed the highest $f_p$ near $60^\circ$N that is not apparent in the CIRS inversions; the hemispheric asymmetry appears more pronounced in the IRIS inversion; and the equatorial minimum in $f_p$ appears more subdued in the IRIS inversion than in the CIRS results.  Note that these quantitative differences are within the $\approx 0.02$ uncertainty on the CIRS and IRIS inversions, and could be neglecting important systematic calibration differences between the two datasets.   The degree of disequilibrium in Fig. \ref{iris_compare}c shows this asymmetry more vividly - sub-equilibrium conditions cover much of the northern hemisphere in the CIRS epoch, whereas super-equilibrium conditions are apparent at $p>200$ mbar from $15-60^\circ$N in the Voyager epoch.  Furthermore, the sub-equilibrium conditions in the equatorial zone do not extend as high during the Voyager observations as during the Cassini observations.  However, the contours in \mbox{Fig. \ref{iris_compare}c} are spaced every 0.01, similar to the uncertainty on $f_p$, so these apparent offsets are may not be particularly significant.  Differences at the equatorial tropopause may be related to inter-annual variability in the oscillating wind and temperature fields above the tropopause, whereas the different hemispheric contrasts are harder to explain.  The differing spectral constraints offered by CIRS and IRIS may also be corrupting the comparison, but we cannot entirely discount a difference in the phasing of the seasonal inter-hemispheric transport between the two epochs (i.e., the strength of the sinking motion at springtime high latitudes).

In summary, although hemispheric contrasts in temperature appear to be broadly reproducible from one Saturn year to the next, hemispheric asymmetries in the para-H$_2$ fraction appear different between the two datasets.  Voyager-1 found the largest $f_p$ at northern high latitudes, and it is notable that Cassini has identified an increasing $f_p$ at the same location as springtime progressed. Equatorial conditions are modulated by a dynamic perturbation that is not locked to the seasonal cycle.  As suggested by the stratospheric results of \citet{14sinclair}, we propose that Saturn's equatorial oscillation has a period that differs from half a Saturn year at the tropopause (i.e., it is not rigorously `semi-annual' and should be referred to as a quasi-periodic oscillation). 

%%%%%%%%%%%%%%%%%%%%%%%%%%%%%%%%%%%%%%%%%%%%%%
%%%%%%%%%%%%%%%%%%%%%%%%%%%%%%%%%%%%%%%%%%%%%%
%%%%%%%%%%%%%%%%%%%%%%%%%%%%%%%%%%%%%%%%%%%%%%
\section{Discussion}
\label{discuss}

Cassini/CIRS provides the capability to measure the zonal mean temperature structure as a function of time, latitude and pressure, from which we can derive several atmospheric diagnostics to understand the dynamic response to seasonal variability.   These include:  (i) the static stability of the atmosphere as a measure of its susceptibility to convective motions; (ii) use of the thermal windshear relation to determine the zonal wind field as a function of altitude, and therefore extending studies of vorticity gradients far above the cloud-tops; and (iii) the importance of these vorticity gradients in confining and guiding planetary wave activity \citep[e.g.][]{96achterberg,06li}.  Given a suitable and accurate radiative-convective model, deviations of the thermal structure from the expectations of radiative balance could be used to estimate the sense and magnitude of vertical motions \citep[e.g.][]{90conrath}, but we leave that to a future study given the difficulties in reproducing Saturn's tropospheric temperature and aerosol response as a function of season \citep{14guerlet}. We explore some of these quantities, derived directly from the thermal field, in the following sections.    Uncertainties in these derived fields use the MC simulation described in Section \ref{results}.

\subsection{Atmospheric Stratification}
\label{stratification}

The seasonal change in the upper troposphere might be expected to alter the stability of the atmosphere to convective motions, by changing the relative altitude of the radiative-convective (R-C) boundary over time.  This can be studied via the frequency of buoyancy oscillations ($N$, known as the \textit{Brunt V\"{a}is\"{a}l\"{a}} frequency):
\begin{equation}
N^2=\frac{g}{T}\left(\pderiv{T}{z} + \frac{g}{c_p}\right)
\end{equation}
Here $g$ is the gravitational acceleration calculated as a function of altitude and latitude accounting for Saturn's oblateness; $c_p$ is the specific heat capacity calculated for the specific temperature and para-H$_2$ fraction at each altitude and latitude; the quantity $\Gamma_a=g/c_p$ is the dry adiabatic lapse rate and $\Gamma=-\partial T/\partial z$ is the lapse rate measured from the reconstructed temperature profiles. $N$ allows us to compare the stratification from year to year directly - statically unstable regions have $N^2<0$ so that $N$ is imaginary; statically stable regions have $N^2>0$ so that $N$ is real and represents a simple harmonic motion of buoyancy oscillations.  

\begin{figure*}
\includegraphics[angle=0,scale=0.8]{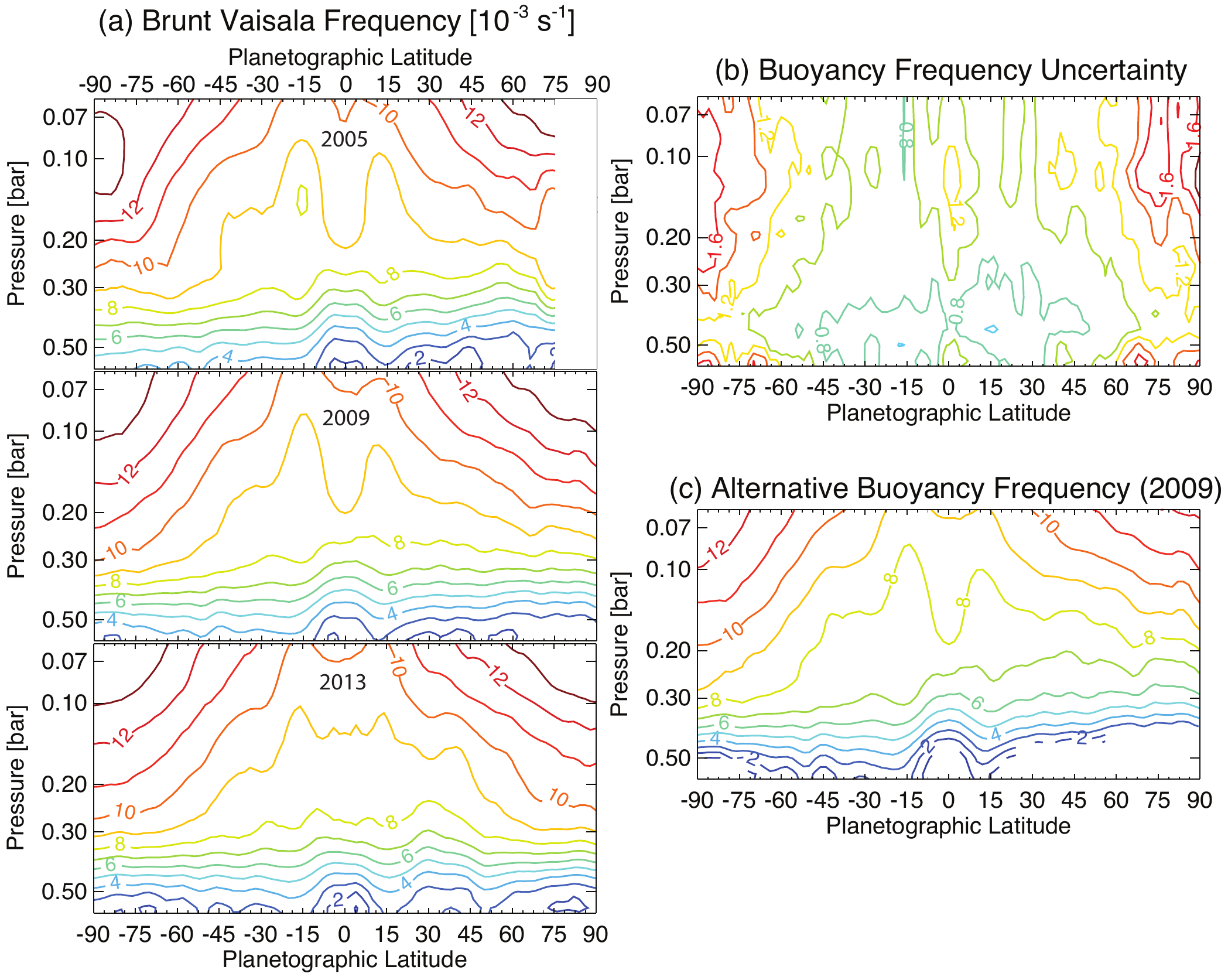}
%\noindent\makebox[\textwidth]{\includegraphics[width=\paperwidth]{figs/buoyancy2.pdf}}
\caption{(a) The \textit{Brunt V\"{a}is\"{a}l\"{a}} buoyancy frequency ($10^{-3}$ s$^{-1}$), a diagnostic of the stratification of Saturn's upper troposphere, as a function of latitude and pressure for Saturn's northern spring equinox $\pm4$ years, calculated from the temperature and gravity field assuming `frozen' equilibrium of ortho-para-H$_2$ as described in the main text.  Spurious oscillations at high north polar latitudes in 2005 are omitted, as these were not directly constrained by CIRS observations.  (b) Uncertainty range on $N$ ($10^{-3}$ s$^{-1}$) in 2009 as a representative example of the time series, calculated via the MC simulation.  (c) Alternative estimate of the buoyancy frequency assuming instantaneous equilibration between the two spin isomers of H$_2$ leading to a generally more unstable troposphere.}
\label{buoyancy}
\end{figure*}

The uncertainty on $N$ can arise from numerous sources - the reconstructed temperature field appears as both a divisor and a derivative in the equation above, and both the temperature and para-H$_2$ fraction determine the specific heat capacity as a function of altitude and latitude.  The MC simulation shows the range of values of $N$ expected from the propagation of uncertainties in $T$ and $f_p$.  The error range on the 2009 values of $N$ (\mbox{Fig. \ref{buoyancy}}) varies from $1\times10^{-3}$ s$^{-1}$ at mid-latitudes and 200 mbar to $1.2\times10^{-3}$ s$^{-1}$ at the equator and $1.6\times10^{-3}$ s$^{-1}$ at the poles.  These error ranges are time variable, with uncertainties exceeding $2\times10^{-3}$ s$^{-1}$ at the poles in 2005.  

In addition, the quantitative values of $c_p$ (and hence $N$) depend on an assumption about the equilibration timescale for ortho-para-H$_2$ conversion.  The standard assumption is to assume the `frozen' limit, where conversion is extremely slow and the two spin isomers can be treated separately - i.e., $c_p$ includes individual contributions of ortho- and para-H$_2$, in addition to contributions from He and CH$_4$.  Latent heat release from the conversion between ortho- and para-H$_2$ is not considered. Given that disequilibrium is detected throughout much of Saturn's troposphere (Fig. \mbox{\ref{parah2_snapshot}}), a long equilibration timescale is a reasonable assumption and will be employed throughout this study.  However, we cannot completely rule out the alternative assumption that equilibration timescales are very short, particularly at higher pressures due to catalysis on the surface of Saturn's tropospheric aerosols \mbox{\citep{82massie}}.  Following the review by \mbox{\citet{09irwin}}, we recalculate the rotational partition function assuming the two extremes:  either an equilibrium mix or treating the two gases separately.   The difference at equinox can be seen in Fig. \mbox{\ref{buoyancy}} - as the equilibrium adiabatic lapse rate $g/c_p$ is smaller than the `frozen' lapse rate, $N$ is systematically smaller by $\approx1\times10^{-3}$ s$^{-1}$ throughout the upper troposphere.  The troposphere is therefore more unstable to buoyancy oscillations than in the frozen case, and $N$ becomes imaginary for $p>500$ mbar (the R-C boundary).   This systematic offset, in addition to the random uncertainties, limits our ability to quantify the buoyancy frequency until some independent estimate of Saturn's static stability is available. 

For the `frozen' assumption, $\Gamma<\Gamma_a$ throughout the 70-600 mbar region in Fig. \ref{buoyancy}, implying that $N^2$ remains positive and the upper troposphere is sub-adiabatic and stably stratified, as expected.  The tropopause has a higher static stability at the equator than at the neighbouring belts, and the stability tends to increase from equator to pole, following similar patterns to those derived from Voyager/IRIS data \citep[Fig. 11 of][which had $N\approx9\times10^{-3}$ s$^{-1}$ at 200 mbar consistent with our Fig. \ref{buoyancy}]{98conrath}.   Small changes are apparent from 2005 to 2013, with the southern hemisphere showing a decreasing $N$ in the 500-mbar region (from $6\times10^{-3}$ s$^{-1}$ to $4\times10^{-3}$ s$^{-1}$), indicating a tendency towards more unstable conditions that is larger than the $1\times10^{-3}$ s$^{-1}$ uncertainty.  The stability becomes smaller at deeper levels and the equator shows some of the smallest values of $N$ near 500 mbar ($3\times10^{-3}$ s$^{-1}$) but this has not changed with time.  In the equilibrium case, the values of $N$ are lower and the R-C boundary is found for $p>500$ mbar.  The tropical region within $\pm30^\circ$ latitude of the equator shows the most substantial changes over time, associated with the warming of the NEB and cooling of the SEB, but these remain sub-adiabatic at all times.  Finally, northern mid-latitudes show a change between the 2009 equinox and the present day that is comparable to our levels of uncertainty, as the warming due to Saturn's northern storm \citep{11fletcher_storm, 14achterberg} decreased $\Gamma$ and therefore also $N$. Unfortunately we cannot determine whether this stratification change preceded the storm eruption (i.e., to relate upper tropospheric stability to the storm triggering) due to the time-sampling of the CIRS data.  These `frozen' estimates of $N$ will be used in the calculation of potential vorticity gradients in the following sections.

%
%The warming and cooling of Saturn's troposphere also affects the scale height, $H=R*T/g$, where $R*$ is the mean molecular weight, $2.3\times10^{-3]$ kg.  $H$ drops from the equator (approximately 34 km at 100 mbar) to the pole (24-27 km) due to the increasing $g$, but also shows seasonal variability.  Although scale height changes between 2005 and 2013 were small ($\Delta H=1-2$ km at 100 mbar), SEB scale heights declined while NEB heights increased, and southern high latitude values of $H$ decreased while northern high latitude $H$ increased.  

\subsection{Tropospheric Winds}
\label{windshear}

If we assume geostrophic balance, then the shifting meridional temperature gradients ($\partial{T}/\partial{y}$, where $y$ is the north-south distance) are related to the vertical shear on zonal jets ($\partial{u}/\partial{z}$) via the thermal windshear equation \citep[e.g.,][p120]{87andrews}.  \citet{09read} used Cassini temperature observations during southern summer \citep{07fletcher_temp} to present a snapshot of the vertical windshear, observing a general decay of the zonal jets with height that was strongest in the southern stratosphere. They called for further measurements of Saturn's wind field throughout the seasonal cycle, so we use the thermal windshear equation in the zonal direction:
\begin{equation}
f\pderiv{u}{\ln \left( p \right)} = \frac{R}{a} \pderiv{T}{\psi} = R\pderiv{T}{y} = -fH\pderiv{u}{z}
\end{equation}
This equation allows us to integrate the zonal winds $u(\psi)$ (where $\psi$ is the latitude) as a function of altitude ($z$), and therefore to explore gradients in the wind fields in the horizontal and vertical directions.  $f=2\Omega \sin(\psi)$ is the Coriolis parameter (the vertical component of the planetary rotation vector, where $\Omega$ is the planetary angular velocity); $a(\psi)$ is the planetary radius; $R$ is the molar gas constant divided by the mean molar weight of Saturn's atmosphere; and $H=RT/g$ is the scale height of the atmosphere.

\begin{figure*}
\begin{centering}
\includegraphics[angle=0,scale=0.7]{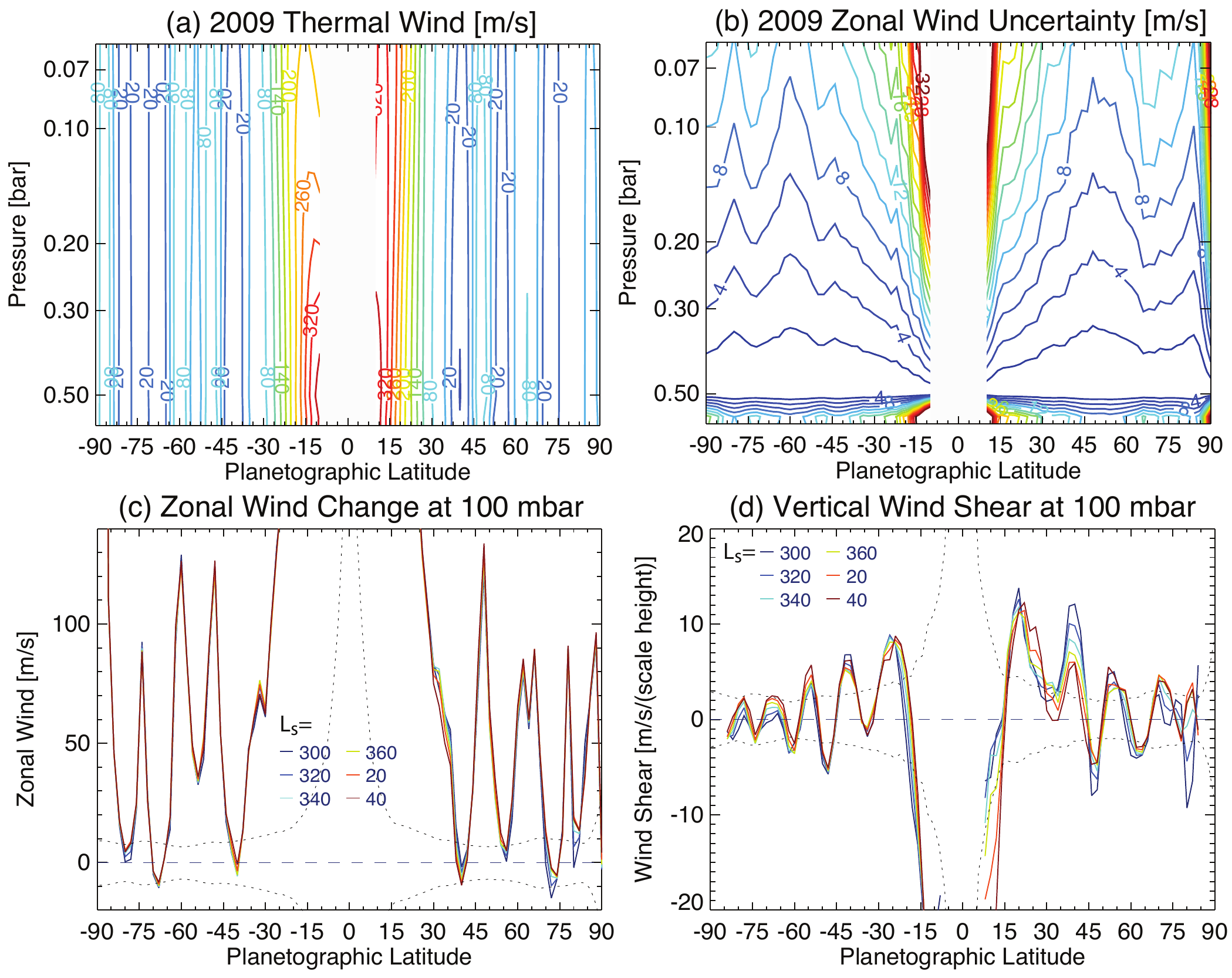}
\caption{Zonal mean winds predicted by integrating the thermal windshear equation using the cloud-top winds of \citet{11garcia}.  Winds are relatively stable, with panel (a) showing the annual-average zonal winds in 2009 (equinox) with a contour spacing of 30 m/s - regions within $\pm10^\circ$ of the equator are omitted because the thermal wind balance becomes invalid at low latitudes.  Panel (b) shows the range of uncertainty on the zonal winds in 2009 from the MC simulation with contours spaced every 2 m/s.  Panel (c) shows the small changes in the zonal wind field at 100 mbar (solid lines for different values of $L_s$) compared to the uncertainty envelope (dotted line).  Panel (d) shows the zonal windshear per scale height at 100 mbar for the same six $L_s$ values, also compared to the uncertainty envelope (dotted line).}
\label{zonalwind}
\end{centering}
\end{figure*}

In this work, the interpolated temperature fields are used to calculate the changing $\partial{u}/\partial{z}$ with time (Fig. \ref{zonalwind}), which is integrated to give $u(z,\psi)$, assuming the wind fields derived from continuum-band Cassini imaging by \citet{11garcia} as a time-invariant boundary condition at 500 mbar.  \citet{11garcia} calculated these velocities using the System III rotational reference frame measured by the Voyager mission \citet{07seidelmann}.  \mbox{Fig. \ref{zonalwind}a} shows a typical latitude-altitude cross section of Saturn's tropospheric winds in 2009, and we find that seasonal variations from this structure are small.  As the thermal wind equation involves spatial gradients of the interpolated temperature fields, the MC technique was used to investigate the range of uncertainties in $u(z,\psi)$, shown in \mbox{Fig. \ref{zonalwind}b}.  Uncertainties are smallest at 500-mbar where we specify the cloud-top winds \citep[although these neglect the 5-10 m/s uncertainties on the cloud tracking measurements,][]{11garcia}, and grow with altitude due to the uncertainties in integration, being of order 10 m/s at the tropopause.  We caution that the absolute magnitude of the zonal winds at each altitude is extremely sensitive to the uncertainty in the location of the cloud-tracked winds, particularly if the altitude of the cloud tracers is latitudinally variable, which would add further systematic error to our measurements.  

Fig. \ref{zonalwind}c shows that changes over the ten-year span of these observations are small at most latitudes, and within the uncertainty derived from the MC simulation.  Variations only appear significant in Saturn's northern hemisphere, particularly for the broad retrograde jets in the hemisphere emerging from winter.  The sharper, narrower prograde jet peaks experience negative shear with altitude and appear largely unaffected over time.  In the southern hemisphere, we see a trend of increasingly positive (prograde) shear on the broad westward jets from summer through to autumn (Fig. \ref{zonalwind}d), but this has limited effect on the zonal winds at 100 mbar (Fig. \ref{zonalwind}c).  In the north, the picture is more complex.  Westward jets at 100 mbar poleward of $45^\circ$N have become more eastward with time by approximately 10 m/s, corresponding to an increasingly positive (prograde) vertical shear over time.  Although this change is equivalent to our uncertainty, the trend with time is continuous rather than random.  The exception is the westward jet at 39$^\circ$N (associated with the outbreak of the 2010 disturbance), which has become more westward with time due to the increasingly retrograde (negative) vertical shear.  This is consistent with the post-storm evolution of the thermal field noted by \citet{14achterberg} and the cloud-tracked wind velocities from \citet{13sayanagi}.  

The classical thermal wind equation relies on an evaluation of the Coriolis parameter, and therefore becomes unreliable at low latitudes, as indicated by the growth of errors in Fig. \ref{zonalwind}.  Nevertheless, the flanks of the equatorial jet show some interesting variability, with the vertical shear at $10^\circ$N becoming increasingly negative (retrograde) over time.  The lack of similar changes at $10^\circ$S suggests a process acting in the northern hemisphere that is absent from the south.  Interestingly, this process is making the zonal windshear profile in Fig. \ref{zonalwind}d appear more symmetric about the equator with time, and this latitude ($10^\circ$N) was identified by \citet{13li} as the location of the largest difference between tropospheric winds derived from Voyager and Cassini data.  This difference between the hemispheres could be caused by a cross-equatorial flow such as those hypothesised by \citet{12friedson} for higher altitudes.  Alternatively, interactions between the zonal-mean flow and waves emitted from the northern storm and `beacon' \citep{12fletcher} might also serve to modify the flow of the equatorial jet on its northern flank.  In summary, while the southern summer/autumn jet system appears to evolve very slowly, the changes to the northern jets are more substantial, possibly as a result of (i) the downward transport at high latitudes suggested by the para-H$_2$ distribution, or (ii) the seasonal heating of the upper troposphere and associated hazes; or (iii) the disruption caused by Saturn's northern storm in 2010-11.  

\subsection{Tropospheric Potential Vorticity}

Having derived Saturn's tropospheric wind field $u(z,\psi)$, we now explore the implications of gradients in these zonal flows on dynamic activity.  Our goal is an understanding of the possibilities for wave propagation in the upper troposphere, but we start by assessing quasi-geostropic potential vorticity (QGPV, $q_G$) gradients as a function of latitude, altitude and time.  QGPV gradients have previously been studied for Jupiter and Saturn \citep{06read_jup,09read}, showing them to be a powerful diagnostic tool for understanding dynamic flows in a stably-stratified atmosphere.  We assume that quasi-geostrophy is valid given the small Rossby numbers associated with Saturn's atmospheric flows (i.e., Rossby numbers smaller than one).  The meridional QGPV gradient \citep[sometimes called `effective beta,' $\beta_e=\partial{q_G}/\partial{y}$, e.g., p127 of][]{87andrews} is assembled from the sum of three terms:  (1) $\beta=\partial{f}/\partial{y}$, the northward gradient of planetary vorticity (i.e., the Coriolis parameter); (2) $\beta_y=-\partial^2u/\partial{y^2}$, the meridional curvature of the zonal wind field; and (3) $\beta_z$, the vertical curvature of the wind field.   This $\beta_e$ is only strictly valid when the meridional length scales are smaller than the planetary radius and longitudinal temperature contrasts are small.  Both approximations hold for the scales of Saturn's belt/zone motions considered here.  Following the definitions of \citet{87andrews, 11sanchez}:
\begin{eqnarray}
\beta_e=\beta+\beta_y+\beta_z \\
\beta_e=\beta-\frac{\partial^2u}{\partial y^2}-\frac{1}{\rho}\pderiv{}{z}\left(\rho \frac{f^2}{N^2}\pderiv{u}{z}\right) 
%\beta_e=\beta-\frac{\partial^2u}{\partial y^2}+g\pderiv{}{p}\left(\rho \frac{f^2}{N^2}\pderiv{u}{z}\right)
\end{eqnarray}
Each variable has been defined previously, with the exception of the atmospheric density $\rho=p/RT$ from the ideal gas equation.  The full two-dimensional (latitude, pressure) variation of each term (gravity, heat capacity, temperature, scale height, density, zonal wind, buoyancy frequency) is accounted for in the calculation.  $\beta_y$ therefore uses the zonal wind field calculated at all altitudes via the thermal windshear relation (Section \ref{windshear}); $\beta_z$ uses the vertical windshear and buoyancy frequency (Section \ref{stratification}). 

% a full beta-plane approximation would have beta and f constant at the value for a reference latitude.

\begin{figure}
\begin{centering}
\includegraphics[angle=0,scale=0.65]{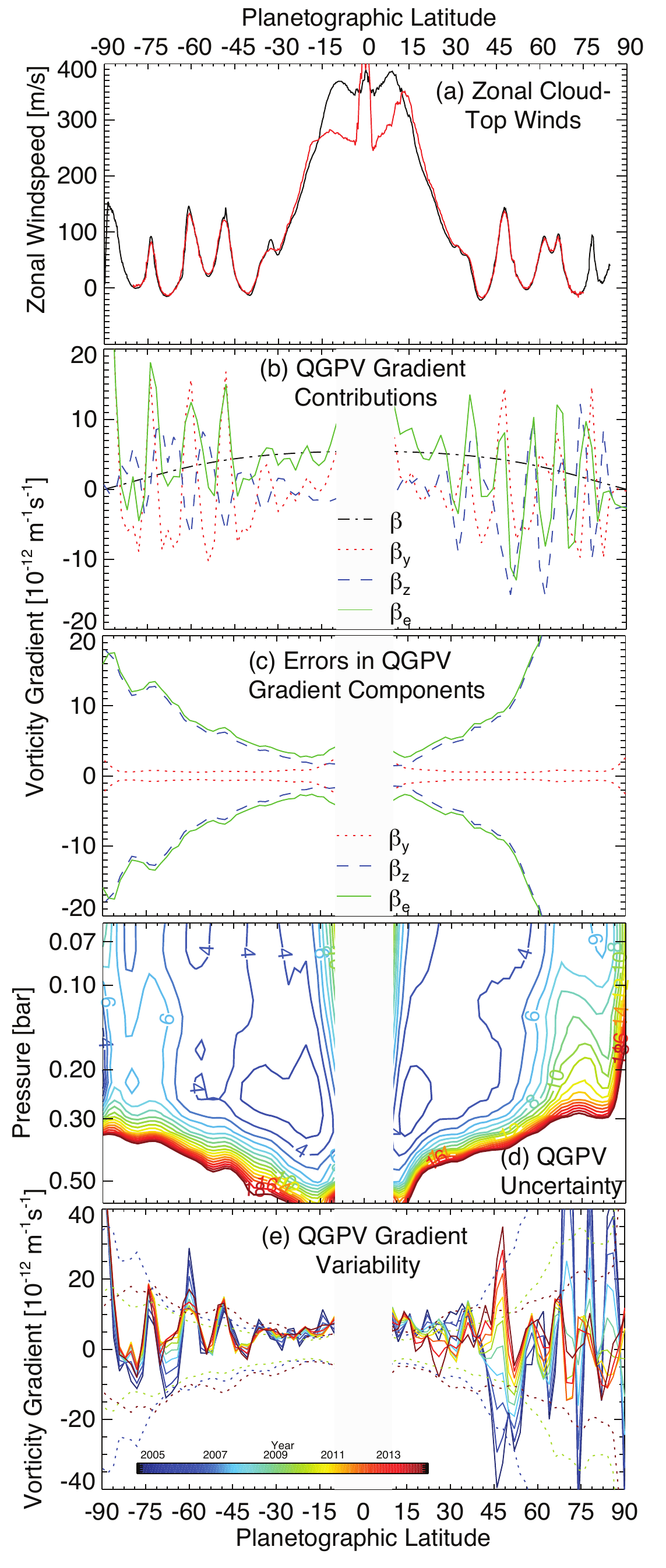}
\caption{Evolution of Saturn's tropospheric quasi-geostrophic potential vorticity (QGPV) gradients as a function of time and latitude.  Panel (a) shows the cloud-top zonal winds from \citet{11garcia}, comparing both continuum-band (black) and methane-band (red) measurements of the wind speeds which sample deep winds ($\approx500$ mbar) and upper tropospheric winds, respectively.  Panel (b) shows the contributions to the `effective beta' ($\beta_e$) QGPV gradient at 330 mbar at Saturn's equinox in 2009, and panel (c) shows the uncertainties in these values from the MC simulation.  Panel (d) shows the variation of $\beta_e$ uncertainty with latitude and pressure, in units of $10^{-12}$ m$^{-1}$s$^{-1}$ - contours higher than $20\times10^{-12}$ m$^{-1}$s$^{-1}$ are not shown.   Panel (e) shows the evolution of $\beta_e$ as a function of year (solid lines with colours shown by the inset key), which causes changes in the QGPV staircase structure \citep{09read} over time. Dotted lines show the time-dependent uncertainty in $\beta_e$ for 2005, 2009 and 2013 from the MC simulation.}
\label{qgpv}
\end{centering}
\end{figure}

The effective $\beta$ and its three components are shown in Fig. \ref{qgpv}(b) as a function of latitude for 2009 at the 330-mbar level (i.e., averaging the reconstructed temperature fields and winds for the year surrounding Saturn's equinox).  Values near the equator are omitted, as the small Coriolis parameter limits the reliability of the extrapolated wind field at low latitudes.  The $\beta_y$ and $\beta_z$ terms introduce a strong variability of the QGPV gradient with latitude, although the two are largely anti-correlated with one another.   Unfortunately, the necessity for taking spatial gradients of the temperatures, winds and para-H$_2$ fields serves to magnify uncertainties, so we use the MC simulation (100 realisations of Gaussian perturbations to the 3000+ $T(p)$ and $f_p(p)$ profiles) to quantify the range of errors on $\beta_y$, $\beta_z$ and $\beta_e$ in \mbox{Fig. \ref{qgpv}c}. As expected, the uncertainty on $\beta_e$ (shown for 2009 in \mbox{Fig. \ref{qgpv}d)} is dominated by the vertical curvature of the wind field $\beta_z$ because of the trend of increasing uncertainty on the zonal wind with height.  In comparison, the meridional curvature of the winds $\beta_y$ provides a smaller contribution to the uncertainty, given that the meridional gradients of uncertainty in the winds in \mbox{Fig. \ref{zonalwind}b} are rather small.  \mbox{Fig. \ref{qgpv}d} also shows where the uncertainty in $\beta_e$ becomes so large ($>20\times10^{-12}$ m$^{-1}$s$^{-1}$) that it dwarfs the meridional changes in $\beta_e$, i.e. for $p>600$ mbar in the mid-latitudes and for $p>300$ mbar at high latitudes.  QGPV gradients are more uncertain in the northern hemisphere due to the sparser time series and cooler winter temperatures (i.e., larger retrieval uncertainties).  

Despite the large uncertainties, Fig. \ref{qgpv}b-c indicate that $\beta_e$ shows reversals in sign as a function of latitude that are larger than the errors, particularly equatorward of $\pm60^\circ$.  QGPV gradients at high latitudes are too uncertain to draw conclusions.  The strongest gradients are associated with the peaks of the prograde jets, as previously discovered by \citet{09read}.  The QGPV gradient reversals result in a `staircase pattern' in the potential vorticity, with steep steps (i.e., high gradients) associated with the prograde jet peaks, and a tendency to flatten in between (i.e., homogenising the QGPV) in association with minima in $u$. \citet{09read} suggested that the presence of strong positive QGPV gradients indicates that the jets are dynamical boundaries at the edges of regions of intense latitudinally-localised activity and mixing of vorticity.  The mixing serves to remove the initial PV gradients (flattening), moving the sharpest gradients to the edges of the band to be associated with a local jump in QGPV across the prograde jets.  

A sign change for the PV gradient with latitude is a necessary (but not sufficient) condition for instability with respect to the Charney-Stern condition \citep{62charney}.  \citet{09read} further point out that Saturn's most interesting dynamic phenomena seem to occur near to where the QGPV gradient changes sign - the polar hexagon, ribbon wave and `storm alley,' and that these are locations where the mean zonal flow appears to violate the neutral-stability criterion relative to Arnol'd's second theorem.  A similar sign reversal is evident in the gradient of absolute vorticity ($\beta-\partial^2u/\partial y^2$) \citep{00sanchez}, but note that a large proportion of the QGPV gradient reversals are driven by the vertical curvature term ($\beta_z$, sometimes referred to as the `stretching' component to distinguish it from the gradient of absolute vorticity), and hence the retrieved temperature field from CIRS is essential for this analysis.  However, a more thorough analysis of the relation between the mean zonal flow and stability violation will likely require numerical modelling of the fluid flows.

The seasonal trends in temperature identified in this work permit an assessment of the variability of the `staircase structure' in QGPV, as shown in Fig. \ref{qgpv}(e).  The QGPV gradients have evolved over the ten years since the analysis of \citet{09read}, but we caution that the uncertainty on $\beta_e$ is time-dependent, being larger at the start and end of the time series than in 2009.  Annual-average uncertainties are depicted as dotted error envelopes in \mbox{Fig. \ref{qgpv}e}, and show that $\beta_e$ changes with time are dwarfed by the uncertainty.  Despite this quantitative uncertainty, qualitative trends are visible:  at southern latitudes, we observe a weakening of the QGPV gradient associated with the $60.9^\circ$S prograde jet, whereas those associated with jets at $73.9^\circ$S and $48.2^\circ$S have remained largely unaltered.  In between the $60.9^\circ$S and $73.9^\circ$S jets, the QGPV gradient has shifted from being very negative to being flatter, and more `step-like.'  Tropical latitudes between $30^\circ$S and $30^\circ$N, dominated by Saturn's broad equatorial flow, have remained largely unaltered over time.  The picture in the northern hemisphere is extremely confusing: uncertainties poleward of $60^\circ$N are so large that the changes are not deemed trustworthy there.  However, the QGPV gradient associated with the prograde jet at $47.8^\circ$N has altered substantially throughout the mission, outside of the uncertainty envelope.  We note that this jet marked the edge of the `warm anomaly' identified at depth in the northern winter hemisphere by \mbox{\citet{07fletcher_temp}} (\mbox{Fig. \ref{temp_snapshot}a}) that disappeared as the mission progressed, and strong QGPV gradients were already noted here by \mbox{\citet{09read}}.  We speculate that the changing sign of the QGPV gradient could have affected the stability properties of the atmosphere in this region prior to the storm eruption in 2010 in the nearby retrograde jet at $39^\circ$N.  Furthermore, the storm eruption near $39^\circ$N could have provided intense vorticity mixing that pushed the sharpest gradients of QGPV northward to the prograde jet at $47.8^\circ$N.  Firm conclusions about the stability properties in Saturn's northern mid-latitudes would require a more focussed study of QGPV gradients surrounding the storm epoch.  

In summary, the meridional gradient of QGPV has been assessed as a function of time, latitude and pressure, and was found to be variable over the ten year span of Cassini's measurements despite the large uncertainties involved in this calculation.  The implications for these gradient changes, both in terms of the vigour of atmospheric mixing and the violation of stability criteria, remain poorly understood, but QGPV derived from temperatures and wind data may prove to be an invaluable tool for studying giant planet meteorology.  Finally, the changing QGPV gradients could potentially act as a baroclinic source of planetary waves \citep{96achterberg}, whose propagation we discuss in the next section.

\subsection{Planetary Wave Activity}

A final application of the CIRS-derived seasonal temperature profiles and associated QGPV gradients is a study of potential wave activity in Saturn's troposphere.  \citet{96achterberg} previously used Voyager/IRIS temperature retrievals to show how the meridional and vertical curvature of the wind fields served to confine the propagation of planetary waves (or Rossby waves) to particular latitude and altitude domains.  Rossby waves owe their existence to the change in the Coriolis force with latitude acting as a restoring force (i.e., $\beta_e$), but the ability to propagate in the vertical depends on the index of refraction of the background medium ($\nu^2$) \citep{61charney}, which can be estimated from the quantities derived above.  The dispersion relation for a 3D Rossby wave in a baroclinic atmosphere (assuming a beta plane) allows us to estimate the zonal phase speed of the wave ($c_x$) relative to the mean zonal wind ($u$) as follows: \citep{11sanchez}:
\begin{equation}\label{eq:dispersion}
u-c_x = \frac{\beta_e}{k^2 + l^2 + (f^2/N^2)(m^2+n^2)}
\end{equation}
where $k$, $l$ and $m$ are the zonal, meridional and vertical wave numbers of the wave, respectively, and $n^2=1/(4H^2)$ following \citet{11sanchez}.  However, this omits higher-order vertical derivatives of the buoyancy frequency \citep{96achterberg}, which are important for Saturn as $N$ (i.e., the static stability) is increasing from near zero in the convective troposphere to large stratospheric values in Fig. \ref{buoyancy}.  We include these additional terms in $n^2$ following \citet{82karoly}:
\begin{eqnarray}
n^2=\frac{1}{4H^2}-\frac{N}{H}\pderiv{N^{-1}}{z}+N\frac{\partial^2{N^{-1}}}{\partial{z^2}}\\
n^2=\frac{1}{4H^2}+\frac{N}{H^2}\pderiv{N^{-1}}{\ln{p}}+\frac{N}{H^2}\frac{\partial^2{N^{-1}}}{\partial{(\ln{p})^2}}
\end{eqnarray}
where the last equality re-expresses the derivatives in terms of pressure.  Following \citet{96achterberg}, we rewrite the dispersion relation into a form where parameters related to the background medium ($u$, $N^2$, $H$, $f$ and $\beta_e$) are separated from the characteristics of the wave:
\begin{eqnarray}
\nu^2 \equiv k^2+l^2+\left(\frac{f^2}{N^2}\right)m^2 = \frac{\beta_e}{u-c_x} - \frac{f^2n^2}{N^2} \\
\nu^2 \equiv \frac{\beta_e}{u-c_x} - \frac{f^2}{H^2N^2}\left(\frac{1}{4}+N\pderiv{N^{-1}}{\ln{p}}+N\frac{\partial^2{N^{-1}}}{\partial{(\ln{p})^2}} \right)
\end{eqnarray}
where $\nu$ is the index of refraction that depends only on the background atmosphere and the phase speed of the wave.  The final equation uses the pressure-coordinate version of $n^2$ to provide a useable expression for $\nu^2$ based on the parameters of the background atmosphere.  If $\nu^2>0$ then the refractive index is real, permitting wave solutions that can propagate.  Alternatively, $\nu^2<0$ implies an imaginary refractive index and evanescent waves that can decay exponentially.  The boundary at $\nu^2=0$ can serve as a trapping barrier, keeping the wave within the region of real refractivity via reflections at this boundary (although there will be leakage through the region of imaginary refractive index).  Note that when $u-c_x=0$, $\nu$ will tend towards infinity - these are known as critical surfaces that could also act to confine the waves, causing wave absorption, wave breaking and interactions with the mean zonal flow \citep{96achterberg}, although some transmission of energy through this critical surface remains possible.  Here the wave solutions become non-linear and unsteady, and can lead to the creation of vortices.  In summary, an estimate of $\nu^2$ based on the properties of the background medium permits a crude exploration of the potential for Rossby wave activity in Saturn's troposphere.

However, this derivation of the index of refraction is a simplification, as it assumes separability of the wave solutions in the zonal, meridional and vertical dimensions.  Here we use the Wentzel-Kramer-Brillouin-Jeffries (WKBJ) approximation \mbox{\citep[Section 4.5 of][]{87andrews}}, whereby the wavelengths are assumed to be much smaller than the length scales over which the background atmospheric medium varies. This will not necessarily be the case in the real atmosphere (particularly for small wavenumbers), and wave solutions will not be sinusoidal as they will vary spatially in the meridional ($l$) and vertical ($m$) directions.  We have implicitly assumed that the wave solution is locally sinusoidal in Eq. {\ref{eq:dispersion}}.  A more comprehensive use of the WKBJ approximation would require the launching of rays in multiple directions to properly compute the meridional and vertical wavenumbers from numerical raytracing.  Nevertheless, the equations above can still provide useful basic insights into the propagation of planetary waves in the middle atmosphere \mbox{\citep[][]{87andrews}}.

\begin{figure*}
\begin{centering}
\includegraphics[angle=0,scale=0.8]{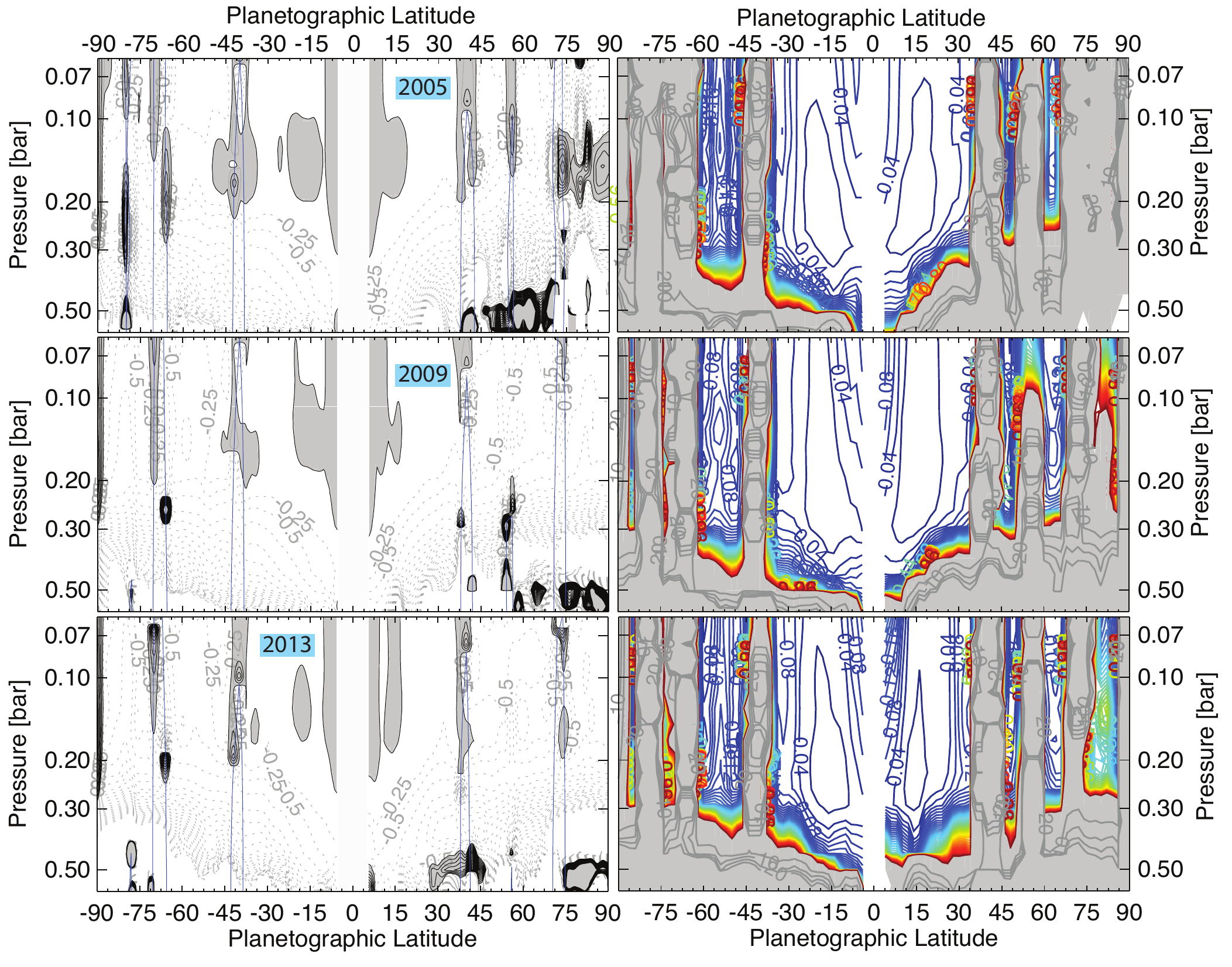}
\caption{\textit{Left column:}  Three examples of annual-average zonal mean contours of the index of refraction ($\nu^2$ in $10^{-12}$ m$^{-2}$) for planetary waves in Saturn's upper troposphere, calculated for the specific case of stationary waves ($c_x=0$).  Contours greater than $\pm1\times10^{-12}$ m$^{-2}$ are shown every $1\times10^{-12}$ m$^{-2}$, with dense contours suggesting rapidly increasing $\nu$ close to critical surfaces.  Contours less than $\pm1\times10^{-12}$ m$^{-2}$ are shown every $0.25\times10^{-12}$ m$^{-2}$.  Regions of real refractive index ($\nu^2>0$) where planetary waves may propagate are shaded grey and feature solid-line contours.  Regions of imaginary refractive index ($\nu^2<0$), where waves are evanescent, are white and feature dotted contours.  Critical surfaces are represented by the blue curves where zonal-mean winds are zero (i.e., where $\nu\rightarrow\infty$).  The equatorial region is omitted due to uncertainties in the wind field.  \textit{Right column:}  Uncertainties in $\nu^2$ ($10^{-12}$ m$^{-2}$) calculated via the MC simulation for the same three years.  Uncertainties smaller than $1\times10^{-12}$ m$^{-2}$ (i.e., the contour spacing in the left column) are shown with contours spaced by $0.02\times10^{-12}$ m$^{-2}$.  Uncertainties larger than $1\times10^{-12}$ m$^{-2}$ are shaded grey and have contours spaced every $5\times10^{-12}$ m$^{-2}$.}
\label{refraction}
\end{centering}
\end{figure*}

Using the two-dimensional fields of $\beta_e$, $N^2$, $H$, $u$ and $f$, we calculate $\nu^2$ as a function of time, latitude and pressure in the left column of Fig. \ref{refraction}.   For simplicity we assume that $c_x=0$ (i.e., phase speeds stationary with respect to the background flow), which implies that critical surfaces only exist for $u\approx0$ m/s in the 2D zonal wind field.  This zero-zonal-velocity surface (i.e., the critical surface where $\nu\rightarrow\infty$ for stationary waves) is included as blue lines in Fig. \ref{refraction}. Regions where $\nu^2>0$ (i.e., a real refractive index permitting wave propagation) are shown in grey, compared to those regions where $\nu^2<0$ (i.e., an imaginary refractive index implying evanescent regions), but we warn the reader that the vertical extent of the positive-index regions will be sensitive to the assumption of $c_x=0$.  Both the critical surfaces and the $\nu^2=0$ boundary appear to change with time in Fig. \ref{refraction}.

The calculated values of $\nu^2$ rely on spatial derivatives of the temperature, wind and composition field (via the calculation of $N^2$), and are therefore subject to large uncertainties that grow extremely rapidly with depth.  The MC simulation permits a crude estimate of the range of expected $\nu^2$ values based on the temperature and para-H$_2$ uncertainties, and annual-average uncertainties are shown in the right-hand column of \mbox{Fig. \ref{refraction}}.  Regions of the figure where uncertainties exceed $1\times10^{-12}$ m$^{-2}$ (the contour spacing on the $\nu^2$ figure) have been shaded to indicate low confidence in the results, and these coincide with regions where the zonal windfield $u$ approaches zero (i.e., the critical surfaces).  As a result, our knowledge of the atmospheric refractivity of Rossby waves is extremely sensitive to uncertainties in Saturn's winds.  The most trustworthy regions are in the tropics, within approximately $\pm35^\circ$ of the equator and for $p<400$ mbar. 

%The full dispersion relation on a sphere (in Mercator coordinates which actually make it simpler than spherical coordiates!) is given in section 2a of Karoly & Hoskins (1982), Journal of the Meterological Society of Japan, vol 60, pg. 109. 

The scale of this uncertainty on $\nu^2$ is rarely reported in the literature, and precludes detailed quantitative analysis.  However, some qualitative trends are apparent that will be worthy of future analysis.  In the specific case of stationary waves, real refractive indices appear to be confined to $p<300$ mbar by the static stability.  These $\nu^2>0$ regions are found surrounding the equator in the 100-200 mbar region with latitudinal boundaries that are relatively stable, but vertical extents than change with time.  For example, the positive-index region near 10-15$^\circ$N extends to lower pressures as time progresses, and the positive-index region near $15-25^\circ$S shrinks in extent over time.  Additional regions of positive $\nu^2$ occur at mid-latitudes in both hemispheres, but here the uncertainties on $\nu^2$ have increased substantially.  The vertical extension of these positive $\nu^2$ regions appears to change from year to year, but trends could be spurious given the level of uncertainty.  We compare our results with Fig. 8 of \citet{96achterberg}, who used raytracing calculations to understand the propagation of quasi-stationary waves based on inversions of Voyager/IRIS spectra.  They also found regions of real refractivity, one equatorward of $25^\circ$N that moved closer to the equator with increasing depth (a similar behaviour is seen in Fig. \ref{refraction}); and a branch of positive refractivity between $30-40^\circ$N that extended down into the upper troposphere, just as we are finding with the Cassini/CIRS data.  They concluded that the propagation of quasi-stationary waves was limited to these latitudes by the curvature of the wind field (i.e., $\beta_y$), and limited to $p<200$ mbar by the static stability (i.e., $\beta_z$), in broad agreement with our findings from Cassini/CIRS.

Adding further confidence to the $\nu^2$ distributions in Fig. \ref{refraction}, the southern mid-latitude region ($35-45^\circ$S) and northern equatorial region ($0-30^\circ$N) have both been identified as locations of strong stratospheric wave activity during the Cassini epoch \citep{14orton_egu}, implying a connection to the refractivity profiles of the upper troposphere.  Furthermore, a near-stationary thermal wave (wavenumber 2) was identified by Voyager IRIS at 130 mbar at northern mid-latitudes during spring \citep{96achterberg}.  In the present analysis, we show that the vertical domain of real refractivity varies because of seasonal shifts in the static stability (e.g., Section \ref{stratification}), whereas the curvature of the wind field is less variable so that the broad meridional structure of the refractive index remains the same over time.  We speculate that slow seasonal changes to the thermal structure, and hence the propagation of planetary waves, influences the variable storm activity with time:  for example, the southern `storm alley' \citep[38.5-46$^\circ$S planetographic,][]{06vasavada} is found to be most active during southern summer/autumn; whereas the eruption of Saturn's northern storm and its influence on the stratosphere \citep{11fischer, 11fletcher_storm} occurred during northern spring.  As wave activity appears to be commonplace in Saturn's stratosphere \citep[e.g.][]{05orton}, and this is where waves are likely to break and deposit their energy \citep{96achterberg}, an extension of this study to lower pressures is required, in addition to proper diagnostics using $k$, $l$, $m$ and $c_x$ for observed planetary waves on Saturn \citep{14orton_egu}.

%%%%%%%%%%%%%%%%%%%%%%%%%%%%%%%%%%%%%%%%%%%%%%
%%%%%%%%%%%%%%%%%%%%%%%%%%%%%%%%%%%%%%%%%%%%%%
%%%%%%%%%%%%%%%%%%%%%%%%%%%%%%%%%%%%%%%%%%%%%%
\section{Conclusions}
\label{conclusion}

Far-infrared Cassini/CIRS spectra of Saturn's hydrogen and helium continuum have been used to track the seasonal evolution of the upper troposphere over a third of a Saturnian year (2004-2014).  Inversions of CIRS spectra allowed us to determine three-dimensional profiles (time, latitude, pressure) of temperatures, para-H$_2$, zonal winds, stratification, potential vorticity gradients and the refractive index for wave propagation.  This builds upon earlier work that captured a snapshot of tropospheric conditions during southern summer \citep{07fletcher_temp}, and allows a quantitative comparison between Voyager (1980) and Cassini (2010) observations at the same point in Saturn's seasonal cycle.  Our conclusions are as follows:

\begin{enumerate}

\item \textbf{Thermal Asymmetries: } Wintertime asymmetries in tropospheric temperatures have largely vanished by late spring, so that temperatures are approximately symmetric about the equator.  The timing of the northern temperature minima and southern temperature maxima varied according to latitude and altitude (occurring later at higher pressures), but summertime maxima occurred sooner after solstice than the wintertime minima, some of which occurred after the spring equinox (2009).  These trends are qualitatively consistent with the expectations of radiative-convective simulations, but quantitative comparisons are hampered by the uncertain heating effects within the upper tropospheric hazes. Saturn's upper troposphere has remained sub-adiabatic and stably-stratified (positive buoyancy frequencies) throughout the period of observations (assuming slow interconversion between ortho- and para-H$_2$), confirming that CIRS senses altitudes above the convective troposphere. Contrasts due to Saturn's belt/zone structure are superimposed onto the seasonal trends, and maxima in $\partial{T}/\partial{y}$ remain co-located with Saturn's cloud-top zonal jets.

\item \textbf{Aerosol Heating: } The thermal perturbation in the 100-250 mbar region due to localised heating in Saturn's haze layer has evolved over time.  The magnitude of the $T(p)$ curvature fell over the southern hemisphere (vanishing poleward of $45^\circ$S by 2014) and grew over the northern hemisphere (starting with the cool zone near $45^\circ$N after 2008) as spring progressed.  By 2014, the NEB and SEB exhibited thermal perturbations of the same relative size.  Unfortunately, CIRS data alone cannot assess whether the shifting magnitude of the perturbation is due simply to enhanced sunlight on a pre-existing and static haze \citep[e.g.][]{12friedson}, or whether the dominant effect is due to the changing colour and opacity of the hazes observed by Cassini \citep[e.g.,][]{12edgington}. 

\item \textbf{Dynamic Heating: }  Anomalously warm regions in the deep troposphere ($p>500$ mbar) evolve over non-seasonal timescales.  These include a warm region between $50-70^\circ$N first identified by \citet{07fletcher_temp}, the heating due to slow subsidence in the aftermath of Saturn's springtime storm at $25-45^\circ$N \citep{14achterberg}, and a similar warm patch identified in Voyager retrievals between $25-45^\circ$N \citep[this work and][]{83conrath}.  These warm regions are at the high-pressure limit of CIRS vertical sensitivity, but could be a dynamical response to the changing stratification of the overlying atmosphere.

\item \textbf{Para-H$_2$: }  Changes in the para-H$_2$ fraction with time are subtle and influenced by dynamics.  $f_p$ showed signs of a general increase in the northern hemisphere (0.02-0.03 at 200-500 mbar) when the strong effects of the low-$f_p$ air lofted by the 2010 storm are discounted.  Northern latitudes poleward of $60^\circ$N show the largest increases near the tropopause, whereas the north polar $f_p$ at 500 mbar was largely unchanged.  This has the effect of changing the symmetric $f_p$ distribution of 2005 into a more asymmetric distribution in 2014 (with higher $f_p$ in the north), suggestive of general subsidence in the springtime hemisphere.  This brings the CIRS 2010 results into line with the Voyager 1980 results, although the latter showed a stronger asymmetry suggesting that para-H$_2$ variability (and therefore the upper tropospheric subsidence) does not proceed at the same rate every Saturnian spring.

\item \textbf{Para-H$_2$ Disequilibrium: }  Disequilibrium in para-H$_2$ reflects primarily the changing temperature structure (and therefore distribution of $f_{p,eqm}$) rather than actual changes in $f_p$ induced by chemical conversion or transport. The reversal of the para-H$_2$ disequilibrium asymmetry, where strong southern super-equilibrium conditions and strong northern sub-equilibrium conditions have both been eroded away, is mostly due to the temperature changes and not to the para-H$_2$ changes described in the previous conclusion.  Neither large-scale interhemispheric transport, nor enhanced catalysis on seasonal hazes, are required to explain this shifting asymmetry.  Nevertheless, dynamics can have a strong influence, with (i) upwelling creating sub-equilibrium conditions at the equator that have declined slightly over time; (ii) upwelling of low-$f_p$ air by the northern storm (which is still in the process of returning the the `normal' conditions at the time of writing); and (iii) increasing subsidence in the spring hemisphere rapidly reducing the sub-equilibrium conditions found there during winter.

\item \textbf{Tropical Oscillations: }  Saturn's upper troposphere in the tropics is influenced by the wave train linked to the stratospheric `semi-annual' oscillation (which should be correctly referred to as a quasi-periodic oscillation).  Equatorial tropopause temperatures have fallen $3\pm1$ K over ten years as a cool anomaly propagated downwards from the stratosphere, and the aerosol heating perturbation appears to have been pushed deeper from 80 to 150 mbar as time progressed.  Temperature, aerosol heating and para-H$_2$ contrasts in the tropics appear markedly different between the Cassini era and the Voyager era (although the general hemispheric asymmetries are well reproduced), suggesting that the influence of the stratospheric oscillation is not in phase with the seasonal cycle at the tropopause (i.e., it is not `semi-annual' at these depths).  Indeed, \citet{14sinclair} also suggested that Voyager and Cassini saw the oscillation at a markedly different phase despite being in the same season.

\item \textbf{Zonal Wind Variability: }  Variable meridional temperature gradients imply that the vertical shear on Saturn's zonal jets changes with time, by values typically smaller than 10 m/s per scale height at the tropopause.  The sharper eastward jets experience negative shear with altitude and are largely unaffected with time; the broader westward jets in the southern hemisphere have experienced increasingly prograde shear with as summer gave way to autumn; and the general picture in the northern hemisphere was complicated by the influence of the springtime storm (the westward jet at 39$^\circ$N has strengthened due to the increasingly retrograde (negative) vertical shear associated with the storm heating).  Intriguingly, the northern flank of the equatorial jet has experienced the largest change in windshear, whereas the southern flank remained largely unaltered, and it is notable that \citet{13li} observed that largest contrasts between Voyager and Cassini-era winds at this latitude ($10^\circ$N).

\item \textbf{Potential Vorticity Variability: }  The meridional quasi-geostrophic potential vorticity (QGPV) gradient (`effective beta'), and hence the `staircase' of potential vorticity identified by \citet{09read}, has varied over the course of the mission.  Tropical latitudes between $30^\circ$S and $30^\circ$N have remained largely unaltered over time, and quantitative conclusions at high latitudes are precluded by the large uncertainties there. Gradients associated with mid-latitude jets have varied, particularly in the northern hemisphere associated with both the storm eruption and the disappearance of a `warm anomaly' at high pressures poleward of $45^\circ$N. These changes in $\beta_e$ have implications for the vigour of atmospheric mixing, and the magnitude of the gradient reversals (i.e., sign changes for $\beta_e$) could imply different degrees of violation of atmospheric stability criteria.  

\item \textbf{Rossby Wave Refractivity: }   Shifting potential vorticity gradients influence the locations of real refractive index for the propagation of planetary waves, and this may modify the confinement regions for wave activity.  The vertical domain of real refractivity is expected to vary because of seasonal shifts in the temperature profile and static stability, whereas the curvature of the wind field is less variable so that the broad meridional structure of the refractive index remains the same over time. We find that stationary Rossby waves would be able to propagate in broad regions near the equator ($20^\circ$N to $20^\circ$S) and $35-45^\circ$ in both hemispheres for $p<300$ mbar (limited by the static stability), consistent with the sparse observations of wave activity in Saturn's upper troposphere and stratosphere.  However, the uncertainties on the refractive index grow extremely large with depth and latitude, so the confinement regions for planetary wave activity remain to be tested by further observations.  

\end{enumerate}

We aim to continue Cassini's tropospheric characterisation beyond 2014, through to Saturn's northern summer solstice ($L_s=90^\circ$) in 2017, and beyond from ground- and space-based observatories.  The implications of the changing wind fields and stratification on the atmospheric stability and the potential for wave propagation remain to be thoroughly explored, and we hope that improved radiative-convective models (fully accounting for aerosol heating) will ultimately allow us to use the deviations from thermal equilibrium to diagnose vertical velocities in the troposphere.  The analysis presented in this paper shows the utility of thermal infrared temperature and para-H$_2$ sounding in revealing the circulation and seasonal variability in a giant planet atmosphere, and the importance of a multi-decadal dataset in understanding the slow variability on these distant worlds.

%%%%%%%%%%%%%%%%%%%%%%%%%%%%%%%%
%%%%%%%%%%%%%%%%%%%%%%%%%%%%%%%%
%%%%%%%%%%%%%%%%%%%%%%%%%%%%%%%%
\section*{Acknowledgments}
The analysis presented in this paper would not have been possible without the tireless efforts of the CIRS operations and calibration team, who were responsible for the design of the imaging sequences, instrument commands and other vital operational tasks.  Fletcher was supported by a Royal Society Research Fellowship at the University of Oxford and the University of Leicester.   The UK authors acknowledge the support of the Science and Technology Facilities Council (STFC).   Orton was supported by grants from NASA to the Jet Propulsion Laboratory, California Institute of Technology.  We thank two anonymous reviewers for their thorough reviews of this manuscript.

\bibliographystyle{elsarticle-harv}
\bibliography{references}




%% \bibitem must have one of the following forms:
%%   \bibitem[Jones et al.(1990)]{key}...
%%   \bibitem[Jones et al.(1990)Jones, Baker, and Williams]{key}...
%%   \bibitem[Jones et al., 1990]{key}...
%%   \bibitem[\protect\citeauthoryear{Jones, Baker, and Williams}{Jones
%%       et al.}{1990}]{key}...
%%   \bibitem[\protect\citeauthoryear{Jones et al.}{1990}]{key}...
%%   \bibitem[\protect\astroncite{Jones et al.}{1990}]{key}...
%%   \bibitem[\protect\citename{Jones et al., }1990]{key}...
%%   \harvarditem[Jones et al.]{Jones, Baker, and Williams}{1990}{key}...
%%


% \end{thebibliography}

\end{document}